\documentclass[12pt]{article}%
\usepackage[hidelinks]{hyperref}
\usepackage{verbatim}
\usepackage{amssymb}
\usepackage{amsthm}
\usepackage{amsfonts}
\usepackage{amsmath}
\usepackage{graphicx}
\usepackage{xcolor}
\usepackage{enumerate, natbib}%

\usepackage{thm-restate}
\usepackage[english]{babel}    % Language: English
\usepackage[ansinew]{inputenc} % Input
\usepackage[T1]{fontenc}       % Font encoding
\usepackage{lmodern,microtype} % Typeface
\usepackage{amsmath,amsthm,amssymb,amsfonts} % AMS math
\usepackage{dsfont,mathrsfs,ushort} % Math style

\usepackage{titlesec,titling}  % Section titles
\usepackage[margin=1in]{geometry}  % Page & margins
\usepackage{setspace,caption}  % Spacing
\usepackage{enumitem,booktabs} % Tables & lists
\usepackage{natbib} % references
\usepackage{pstricks,pst-all}   % Figures
\usepackage{pst-plot,pst-node,pst-3dplot}
\usepackage{sgamevar}           % Strategic form games
\usepackage{hyperref,breakurl}
\usepackage{pgfplots}
\usepackage{longtable}
\usepackage{algorithmic}
\setenumerate{label=\small(\roman*)}
\hypersetup{colorlinks=true,pdfnewwindow=true,pdfstartview=FitH,%
	pdftitle={A Rationalization of the Weak Axiom of Revealed Preference},pdfauthor={Victor H. Aguiar; Per Hjertstrand; Roberto Serrano; Ozgur Evren},%
	urlcolor=blue!90!red!45!black,citecolor=blue!90!red!45!black,linkcolor=red!90!black}
\DeclareCaptionFont{fancy}{\bfseries\sffamily}
\allowdisplaybreaks

\providecommand{\psreset}{\psset{%
		linewidth=0.3pt,linestyle=solid,linecolor=black,
		dotsize=2.5pt,dotsep=2.5pt,arrowsize=4pt,
		fillstyle=none,fillcolor=white,
		showpoints=false,arrows=-,linearc=0,framearc=0,
		hatchsep=2pt,hatchwidth=0.2pt,nodesep=4pt,opacity=1}
	\psset{gridcolor=black!60, subgridcolor=black!30}
}

\psreset

\usepackage{graphicx}
\graphicspath{ {figures/} }

\titleformat{\section}[block]{\centering\large\bfseries\sffamily}{\thesection.}{0.5em}{}
\titleformat{\subsection}[block]{\flushleft\bfseries}{\thesubsection.}{0.5em}{}
\titleformat{\subsection}[block]{\flushleft\bfseries\sffamily}{\thesubsection.}{0.5em}{}
\titleformat{\subsubsection}[runin]{\normalsize\itshape}{\bfseries\upshape\sffamily\thesubsubsection.}{0.5em}{}[.--\:]
\renewcommand{\thesubsubsection}{\arabic{section}.\arabic{subsection}.\alph{subsubsection}}
\titlespacing{\section}{0ex}{10ex}{5ex}
\titlespacing{\subsection}{0in}{6ex}{3ex}
\titlespacing{\subsubsection}{0mm}{2ex}{0.5em}
\pretitle{\begin{center}\LARGE\bfseries\sffamily}
	\posttitle{\par\end{center}\vskip 0.5em}
\preauthor{\begin{center} \large \lineskip 0.5em\begin{tabular}[t]{c}}
		\postauthor{\end{tabular}\par\end{center}}
\predate{\begin{center}\small}
	\postdate{\par\end{center}}
\providecommand{\abstitle}[1]{{\par\vspace*{2ex}\small\bfseries\sffamily #1}\hspace*{1ex}}
\renewenvironment{abstract}%
{\begin{center}\begin{minipage}{0.8\linewidth}%
			\setlength{\parindent}{0.0em}\abstitle{Abstract}\small}%
		{\end{minipage}\end{center}\vfill\clearpage}

\providecommand{\U}[1]{\protect\rule{.1in}{.1in}}
\newtheorem{theorem}{Theorem}
\usepackage{filecontents}

\newtheorem{axiom}{Axiom}

\newtheorem{claim}{Claim}

\newtheorem{definition}{Definition}
\newtheorem{example}{Example}
\newtheorem*{example*}{Example}

\newtheorem{lemma}{Lemma}

\makeatletter
\def\@biblabel#1{\hspace*{-\labelsep}}
\makeatother

\pgfplotsset{compat=1.15}
\usepackage{tikz}
\usetikzlibrary{matrix,chains,positioning,decorations.pathreplacing,arrows}

\usetikzlibrary{matrix,arrows.meta,positioning}

\usepackage{animate}
\begin{document}
\title{A Rationalization of the Weak Axiom of Revealed Preference\thanks{\scriptsize This paper supersedes our previous draft titled \textquotedblleft The
Theory of Weak Revealed Preference.\textquotedblright{}  We are grateful
to Roy Allen, Tilman B{\"o}rgers, Peter Caradonna, Chris Chambers, Pawel Dziewulski, Federico Echenique,  Francoise Forges, Mira Frick, Amanda Friedenberg, Charles Gauthier, Maria Goltsman, Reinhard
John, Nail Kashaev, Bart Lipman, Andreu Mas-Colell, Yusufcan Masatlioglu, Rosa Matzkin, Hiroki Nishimura, Efe Ok,  John Rehbeck, Phil Reny, Marciano Siniscalchi, Al Slivinsky,  and Gerelt
Tserenjigmid for useful comments and encouragement. We thank Judith Levi for her great
editing work. Serrano thanks Universidad Carlos III in Madrid for its warm hospitality and Fundacion ONCE for research support.}}
\author{Victor H. Aguiar\thanks{Department of Economics, Simon Fraser University, British Columbia, Canada, vaguiar@sfu.ca.} \and Per Hjertstrand\thanks{Research Institute of Industrial Economics (IFN), Stockholm, Sweden,
per.hjertstrand@ifn.se. } \and Roberto Serrano\thanks{Department of Economics, Brown University, Providence, Rhode Island,
U.S.A., roberto\_serrano@brown.edu.} \and Ozgur Evren\thanks{New Economic School, Moscow, Russian Federation, oevren@nes.ru.}}
\date{First version: January 2020. This version: November 2025}
\maketitle
\begin{abstract}
\noindent 
\footnotesize
Given a finite collection of choices over budgets, rational consumer behavior is identified with the generalized axiom of revealed preference (GARP).  As an alternative, the weak generalized axiom of revealed preference--WGARP--only rules out strict choice cycles of size two.   While GARP failures are routinely found in both the field and the lab, the same is not true for WGARP.  We study how WGARP-consistent choices can be rationalized as a form of optimal behavior. Our main result is an analogue of the celebrated Afriat's theorem, but for WGARP. We show that choices are consistent with WGARP if and only if they can be rationalized by an asymmetric and strictly monotonic preference function. Equivalently, they can be rationalized by a preference function that admits a coalitional
multi-utility (CMU) representation
with a coherence restriction. 
 A coherent CMU representation aggregates multiple utility functions --selves-- within the individual, so that binary preference reversals do not occur,  while longer preference cycles remain possible. As a result, the class of incomplete, intransitive, and nonconvex preferences compatible with WGARP is studied.
\bigskip{}
\end{abstract}

\section{Introduction}
One of the cornerstones of classical consumer theory is \citeauthor{afriat_construction_1967}'s \citeyearpar{afriat_construction_1967} theorem. It states that a finite collection of bundles chosen from budgets given by prices and income (a dataset) can be rationalized by a continuous, strictly increasing, and concave utility function if and only if the dataset satisfies a consistency condition, called the Generalized Axiom of Revealed Preference (GARP). In revealed-preference theory, one can consider a far weaker requirement, namely, consistent choices over only pairs of observations, as captured by \citeauthor{samuelson_note_1938}'s (\citeyear{samuelson_note_1938}) well-known weak axiom of revealed preference (WARP), or its generalized version for correspondences, the weak generalized axiom of revealed preference (WGARP, originally introduced by \cite{kihlstrom_demand_1976-1}). Remarkably, a rationalization of WARP/WGARP has remained as a long-standing open question.

There are good reasons to pay attention to WGARP. Heuristically, it would appear that GARP should be much harder to get than WGARP as a description of behavior. Indeed, the former, equivalent to rational behavior, implies the symmetry of the Slutsky matrix (\emph{equalities} of all substitution effects for any pair of goods), whereas the latter, as detailed below, is checking a number of simple \emph{inequalities}. Also, evidence obtained from recent empirical and experimental datasets support WGARP but not GARP. For instance, \citet{youmbi2024nonparametricanalysisrandomutility} provides evidence, from the UK family expenditure survey, that choices from a population of consumers are generally inconsistent with GARP, but they do comply with WGARP assuming i.i.d. random preferences. \citet{aguiar2017slutsky} showed that the extent of GARP violations in consumer behavior surpasses that associated with departures from WGARP, using the experimental dataset of \citet{ahn2014estimating}. In a discrete-choice experiment, \citet{Costa-Gomes2022} finds that a significant fraction of subjects satisfy WARP, but not GARP. Allowing for additive stochastic noise, \citet{alos2022identifying} provides experimental evidence against transitive preferences (at the heart of GARP) and in favor of asymmetric and nontransitive preferences, which are consistent with WGARP (see also the classic paper by \cite{Tversky1969}).
\par
Yet, a complete characterization of WGARP is not available. The purpose of this paper is to investigate how Afriat's rationalization notion can be modified to characterize WGARP. That is, our question is  whether WGARP-consistent choices can be rationalized as a form of optimal behavior. Since WGARP allows cyclic choices, such behavior cannot simply be described by maximization of a standard utility function. We therefore study the restrictions governing WGARP-compatible preferences, with a particular focus on the functional representation of such preferences.
\par
To understand WGARP, imagine a consumer in a store facing two bundles of goods, $x^t$ and $x^s$. If the consumer chooses $x^t$ despite it being more expensive than $x^s$, we say she has a \textit{direct revealed preference} for $x^t$ over $x^s$, with strict preference revelation when $x^t$ is strictly more expensive. The consumer is said to have an \textit{indirect revealed preference} for $x^t$ over $x^s$ when there is a chain of direct preference revelations with $x^t$ as its beginning and $x^s$ as its end point.
WGARP mandates that strict revealed-preference cycles of size 2 be absent, i.e., there is no pair of observations such that $x^t$ is directly revealed preferred to $x^s$, and, at the same time, $x^s$ is directly and strictly revealed preferred to $x^t$. GARP adds transitivity to WGARP, requiring the absence of strict revealed-preference cycles of any size, i.e., that there is no pair of observations such that $x^t$ is indirectly revealed preferred to $x^s$ and, at the same time, $x^s$ is directly and strictly revealed preferred to $x^t$.
\par
In order to provide an exact analogue of the classical Afriat Theorem but for WGARP, we work with preference functions, which allow us to bypass the transitivity condition thanks to their flexible structure. Specifically, a preference function $r$ assigns a real number $r(x,y)$ to each pair of bundles $x,y$, with the interpretation that $x$ is preferred to $y$ whenever $r(x,y)\geq 0$. (The classical theory is the special case with $r(x,y)\equiv u(x)-u(y)$ for some utility function $u$.) 

Our main result provides: (i) a minimal rationalization of WGARP in terms of a preference function with two basic properties, namely, asymmetry and local nonsatiation/monotonicity; (ii) a characterization of WGARP in terms of Afriat and Varian inequalities (\cite{varian_nonparametric_1982}) that are useful for empirical work; and (iii) a more structured rationalization by a Coalitional Multi-utility (CMU) function, first introduced in \citet{nishimura2016utility}, that features a number of additional properties. The CMU can be understood as a way to aggregate the preferences of different ``selves'' within the individual; each ``self'' expressing a specific attribute that may be relevant in only \emph{some} pairwise bundle comparisons.  

Using our main result, we find that WGARP-compatible behavior can be associated with a host of incomplete, intransitive, and even nonconvex preference relations. Simple and super majority voting rules as a method of aggregating the preferences of different selves comprise a particularly interesting class of WGARP-compatible models. We also show that a classic CMU model routinely generates violations of WGARP, highlighting the specialty of WGARP-compatible CMU functions that we construct. Overall, we identify a rich but structured class of behavioral models that comply with WGARP.

\par
The paper unfolds as follows. In Section~\ref{sec:prelim} we introduce the necessary preliminaries on revealed preference and the general notion of preference functions, setting the stage for our new rationalizations. Section~\ref{sec:main} presents our main Afriat-type result for WGARP, showing that observed choices satisfy WGARP if and only if they can be rationalized by an asymmetric and strictly increasing preference function---equivalently, by a coherent  CMU representation. In Section~\ref{sec:negative}, we document several negative results relating to WGARP, in particular, the possibility of empty demand correspondences and the compatibility with nonconvex preferences. In Section~\ref{sec:kacyclic}, we define $k$-acyclicity, which rules out strict revealed-preference cycles of size $k$, and connect it to restrictions of the CMU. This gives us conditions ensuring the possibility of counterfactual predictions of demand. Section~\ref{sec:textbook} develops duality results and first-order conditions for optimization problems associated with the CMU preferences. Finally, Section~\ref{sec:conclusion} concludes. Further considerations on empty demand and nonconvex preferences, as well as omitted proofs are collected in Appendices \ref{appendixA} and \ref{app:proofs}, respectively. Extensions (WARP, more abstract settings, and ordinal content of our CMU) are the subject of the Online Appendix.

\section{Preliminaries\label{sec:prelim}}

Suppose that a consumer chooses bundles
consisting of $L\geq2$ goods. There are $T < \infty$  observations, indexed as $t=1,\ldots,T$, each specifying the prices and quantities of the $L$ goods selected by the consumer. The set $\mathbb{T}\equiv\left\{ 1,\ldots,T\right\}$ collects the indices. For every observation $t$, the associated price vector is denoted by $p^t$ and the quantity bundle by $x^t $. The $T$ observations of $(p^t, x^t)$ form the dataset $O^{T} = \{(p^t, x^t)\}_{t\in \mathbb{T}}$.\footnote{For any $x,y\in\mathbb{R}^{L}$, we denote by $xy$ the inner product;
$x\geq y$ means $x_{l}\geq y_{l}$ for all $l=1,\ldots,L$; we write $x>y$
if $x\geq y$ and $x\neq y$, while $x\gg y$ means $x_{l}>y_{l}$ for all
$l=1,\ldots,L$. Further, we set $\mathbb{R}_{+}^{L}=\{x\in\mathbb{R}^{L}:x\geq (0,\ldots,0)\}$
and $\mathbb{R}_{++}^{L}=\{x\in\mathbb{R}^{L}:x\gg(0,\ldots,0)\}$.}

Throughout, we study rationalizability of a dataset by a possibly nonstandard preference relation defined on the space of all possible consumption bundles, $X=\mathbb{R}_{+}^{L}.$ We assume that each observed bundle $x^{t}$ is a nonzero element of $X$, the price vector $p^{t}$ is strictly positive, and  the observations are consistent with Walras' law, that is, the observed expenditure $w^t\equiv p^{t}x^{t}$ equals the consumer's wealth. The set of all possible price-wealth pairs is $P\times W = \mathbb{R}_{++}^{L+1}$. A pair $(p,w)\in P\times W$ induces the budget set $B(p,w)=\{x\in X:px\leq w\}$.

\subsection{Revealed Preference}
\label{RPdefs}

We begin by recalling some key definitions in the revealed-preference
literature, e.g. \citet{varian_nonparametric_1982}.

\begin{definition}
 (Direct revealed preference)
  We say that
$x^{t}$ is directly revealed preferred to $x^{s}$, written $x^{t}\succeq^{R,D}x^{s}$,
when $p^{t}x^{t}\geq p^{t}x^{s}$. Also, $x^{t}$ is strictly and
directly revealed preferred to $x^{s}$, written $x^{t}\succ^{R,D}x^{s}$,
when $p^{t}x^{t}>p^{t}x^{s}$.
\end{definition}

If $x^{t}$ is directly revealed preferred to $x^{s}$,  the consumer chose $x^{t}$ and not $x^{s}$, when both bundles
were available in the budget set $B(p^t,w^t)$. If, in addition, the strict inequality $p^tx^t > p^t x^s$ holds, it becomes possible to find another bundle $y$ in $B(p^t,w^t)$ such that $y\gg x^s$. We then conclude that $x^t$ is directly revealed preferred to the bundle $y$, which happens to be strictly greater than $x^s$. This pattern may be viewed as evidence that the consumer strictly prefers $x^{t}$ to $x^s$.

The definitions above focus on instances that provide direct evidence about the preference ranking of two bundles. If the consumer is equipped with a transitive preference relation, indirect inferences also become possible, along the following lines:

\begin{definition}
 (Indirect revealed preference)
  We say that $x$
is (indirectly) revealed preferred to $y$, written $x\succeq^{R}y$,
when the dataset $O^T$ contains a chain $(x^{t_{1}},x^{t_{2}},\ldots,x^{t_{n}})$ such that $x=x^{t_{1}}\succeq^{R,D}x^{t_{2}}\succeq^{R,D}\cdots\succeq^{R,D}x^{t_{n}}=y$.
Also, $x$ is strictly revealed preferred to $y$, written
$x\succ^{R}y$, when there exists a chain as above and with at least one strict direct preference relation, $\succ^{R,D}$.
\end{definition}

This indirect method allows one to define the consumer's preferences consistently if the dataset satisfies the following property:

\begin{axiom}
 (GARP)
  The generalized axiom of revealed preference
(GARP) holds if there is no pair of observations $s,t\in\mathbb{T}$
such that $x^{t}\succeq^{R}x^{s}$ and $x^{s}\succ^{R,D}x^{t}$.
\end{axiom}

\citet{varian_nonparametric_1982} introduced GARP, and showed that it is equivalent to \citet{afriat_construction_1967}'s Cyclical Consistency axiom. These axioms are compatible with set-valued demand correspondences, as opposed to \citet{Houthakker1950}'s strong axiom of revealed preference (SARP) for single-valued demand functions.

Cyclic or inconsistent choices that violate GARP may stem from a variety of reasons such as indecision, i.e., the absence of well-defined preferences (cf. \citet{Costa-Gomes2022}), imprecise assessment of bundles as in \citet{Luc1956}'s theory of imperfect discrimination, or conflicting objectives of household members, as in the classic Condorcet paradox associated with the majority voting rule. Indeed, the empirical literature mentioned in the Introduction shows that cyclic choice behavior is ubiquitous. 

In this paper, we focus on the following generalization of GARP: 

\begin{axiom}
 (WGARP)
  The weak generalized axiom of revealed preference
(WGARP) holds if there is no pair of observations $s,t\in\mathbb{T}$
such that $x^{t}\succeq^{R,D}x^{s}$ and $x^{s}\succ^{R,D}x^{t}$.
\end{axiom}

This axiom rules out direct revealed-preference cycles of length $2$, as opposed to cycles of arbitrary length considered in GARP. It demands only pairwise observations be consistent, so that we can define the consumer's binary preferences in a meaningful way, while retaining the basic premise that $\succ^{R,D}$ is a subset of the consumer's strict preference relation. 
WGARP and its analogue for single valued demand functions, WARP,\footnote{We leave a formal discussion of single-valued demand functions to the Online Appendix.} have attracted attention since the seminal work of \citet{samuelson_note_1938}, presumably because of their intuitive appeal.  Yet, we will see that WGARP provides substantive behavioral restrictions, and some prominent theories of intransitive behavior are incompatible with it.

When $L=2$, GARP is known to be equivalent to WGARP (\citealt{Banerjee2006}), but this is an exceptional case, as illustrated by the following example:

\begin{example}
 \label{ex:WGARPnotGARP} (\citealt[Example 1, p.467]{keiding_revealed_2013})
 Suppose $L=3$, and there are three observations with prices $p^{1}=(4,1,5)$, $p^{2}=(5,4,1)$,
$p^{3}=(1,5,4)$, and bundles $x^{1}=(4,1,1)$, $x^{2}=(1,4,1)$,
$x^{3}=(1,1,4)$. Note that 
\[p^1 x^1=p^2 x^2=p^3 x^3=22,\quad p^1 x^2=p^2 x^3=p^3 x^1=13,\quad p^1 x^3=p^2 x^1=p^3 x^2=25. 
\]

This yields a strict preference cycle $x^1\succ^{R,D}x^2\succ^{R,D}x^3\succ^{R,D}x^1$. GARP is violated by this cycle of length $3$. But the dataset does not contain a binary cycle, that is, whenever $p^t x^t>p^t x^s$, we also have $p^s x^t >p^s x^s$, consistent with $x^t$ being preferred to $x^s$. This makes the observations compatible with WGARP.

\end{example}

Violations of WGARP are often accompanied by violations of the compensated law of demand. In the next example, the demand for good $2$ increases in response to an increase in its price, despite the fact that the original demand vector remains affordable after the price change.

\begin{example}\label{exp:CLD} %(\citet{aguiar2021cardinal}) 
Consider the dataset $O^2$ with prices $p^1=(1,2)$, $p^2=(1,3)$, and bundles $x^1=(6,0)$, $x^2=(0,2)$. This violates WGARP since $x^2\succeq^{R,D}x^1$ and $x^1\succ^{R,D}x^2$. That is, $x^1$ is affordable when $x^2$ is chosen, but $x^2$ is cheaper than $x^1$ when $x^{1}$ is chosen. While the former statement would be consistent with $x^1$ and $x^2$ being indifferent, the latter statement falsifies that conclusion.
\end{example}

The standard rationalization notion is based on a utility function:

 \begin{definition}
  (Utility rationalization)
   \label{UtilityRat}
   A function $u:X\mapsto\mathbb{R}$ rationalizes a dataset $O^{T}=\{(p^{t},x^{t})\}_{t\in\mathbb{T}}$ if $u(x^{t})\geq u(y)$ for every $y\in B(p^t,w^t)$ and $t\in\mathbb{T}$.
\end{definition}
\citeauthor{afriat_construction_1967}'s (\citeyear{afriat_construction_1967})
fundamental theorem provides several equivalent characterizations of a rationalizable dataset. \bigskip{}
 \\
 \textbf{Theorem.}
{\em  (Afriat's theorem, \cite{varian_nonparametric_1982})
Consider a finite dataset $O^{T}=\{(p^{t},x^{t})\}_{t\in\mathbb{T}}$.
The following statements are equivalent:
\begin{enumerate}
\item The dataset $O^{T}$ can be rationalized by a strictly increasing,
continuous, and concave function $u:X\mapsto\mathbb{R}.$
\item The dataset $O^{T}$ satisfies GARP.
\item There exist numbers $U^{t}$ and $\lambda^{t}>0$ for $t\in\mathbb{T}$
such that the Afriat inequalities:
\[
U^{t}-U^{s}\geq\lambda^{t}p^{t}(x^{t}-x^{s}),
\]
hold for all $s,t\in\mathbb{T}$.
\item There exist numbers $V^{t}$ for $t\in\mathbb{T}$ such that the
Varian inequalities:
\begin{eqnarray*}
\text{if }p^{t}(x^{t}-x^{s}) & \geq & 0,\text{ then }V^{t}-V^{s}\geq0,\\
\text{if }p^{t}(x^{t}-x^{s}) & > & 0,\text{ then }V^{t}-V^{s}>0,
\end{eqnarray*}
hold for all $s,t\in\mathbb{T}$.
\end{enumerate}
}

Statements
(ii), (iii), and (iv) are testable conditions, easily implemented in practice. The theorem shows that these conditions are all equivalent to rationalization by a strictly increasing utility function, which may as well be assumed continuous and concave, without loss of generality.\footnote{In fact, the conclusion of the theorem remains valid if we demand the function $u$ be only locally nonsatiated, a property that is weaker than strict monotonicity. GARP becomes a necessary consequence of rationalization only under such desirability conditions.} \citet{varian_nonparametric_1982} shows that the numbers $U^{t}$
and $\lambda^{t}$ in statement (iii) can be interpreted as measures
of the utility and marginal utility of income, respectively, at observation
$t$. Analogously, the
numbers $V^{t}$ in statement (iv) can be interpreted as measures
of the utility levels at the observed demand vectors (\cite{Demuynck2019}).

How can we rationalize a dataset that satisfies WGARP? Rationalization by a strictly increasing (or locally nonsatiated) utility function will no longer be possible if the dataset violates GARP, as in Example~\ref{ex:WGARPnotGARP}. The main finding of this paper is a theorem that characterizes WGARP using a more general functional form. Analogously to Afriat's theorem, we will demand continuity and strict monotonicity conditions, and also some novel functional restrictions suitable for our general approach.

\subsection{Preference Functions}
\label{subsec:preffunctions}

The standard consumer theory presumes a utility function $u:X\mapsto\mathbb{R}$ such that $u(x)-u(y)\geq 0$ if and only if the bundle $x$ is preferred to the bundle $y$. We represent the consumer's preferences with the following more general form:

\begin{definition} 
(Preference function) 
A preference function $r:X\times X\to\mathbb{R}$ maps ordered pairs of commodity bundles to real numbers: for each pair of bundles $(x,y)$, it assigns the real number $r(x,y)$. 
\end{definition}

A preference function $r$ induces a binary relation $\succeq_{r}$ on $X$ by the rule $x\succeq_r y$ if and only if $r(x,y)\geq 0$. Note that the preference relation $\succeq_{r}$ may be intransitive and incomplete. The symmetric part $\sim_{r}$ of this relation and its strict part $\succ_{r}$ are defined as usual: $x\sim_{r} y$ if and only if $x\succeq_{r} y$ and $y\succeq_{r} x$, while $x\succ_{r} y$ if and only if $x\succeq_{r} y$ and $\neg (y\succeq_{r} x).$ We say that $x$ is \emph{preferred} to $y$ whenever $x\succeq_{r} y$. Indifference and strict preference are defined analogously. When we say $r$ is reflexive, transitive, complete, intransitive, or incomplete, we mean that $\succeq_{r}$ has the corresponding property.

The following definition introduces three important properties of preference functions that feature in our characterizations of WGARP.

 \begin{definition}
  \label{properties}
   Consider a preference
function $r:X\times X\rightarrow\mathbb{R}$. We say that:
\begin{enumerate}

\item $r$ is \emph{strictly increasing} if for all $x,y,z\in X$, $x> z$
implies $r(x,y)>r(z,y)$.

\item $r$ is \emph{asymmetric} if for all $x,y\in X$, $r(x,y)\geq(>)0$
implies $r(y,x)\leq(<)0$.

\item $r$ is \emph{skew-symmetric}
if for all $x,y\in X$, we have $r(x,y)=-r(y,x)$.
%\item $r$ is \emph{quasiconcave} if for all $x,y,z\in X$ and  $\lambda\in (0,1)$,
%we have $r(\lambda x+(1-\lambda) z,y)\geq\min\left\{ r(x,y),r(z,y)\right\}.$
%and \emph{strictly quasiconcave} if, for any $0<\lambda<1$, the inequality
%is strict whenever $x\neq z$.

%\item $r$ is \emph{concave} if for all $x,y,z\in X$ and $\lambda\in (0,1)$,
%we have $r(\lambda x+(1-\lambda) z,y)\geq\lambda r(x,y)+(1-\lambda)r(z,y)$.
%and \emph{strictly concave} if, for any $0<\lambda<1$, the inequality
%is strict whenever $x\neq z$.

%\item $r$ is \emph{piecewise concave} if there is a nonempty and finite set $\mathbb{K}$   of concave
%functions in the first argument, $f_{t}\left(x,y\right)$,  $f_t\in\mathbb{K}$,   
 %such that $r\left(x,y\right)=\max_{f_t\in\mathbb{K}}\left\{ f_{t}\left(x,y\right)\right\} %$.\footnote{See \citet{Zangwill1967} and \citet{tsevendorj2001piecewise} for a detailed discussion of %piecewise-concave functions.}
%A preference function $r$ is \emph{strictly piecewise concave} if there is a similar set
%of strictly concave functions.
\end{enumerate} 
\end{definition}

The strict monotonicity of $r$, as defined in statement (i) above, is logically distinct from the condition that $x> z$
implies $x\succ_r z$. Rather, when $r$ is strictly increasing, the associated preference relation $\succeq_r$ becomes consistent with the ``greater than'' relation $>$, in the sense that $x> z$ and $z\succeq_r y$ imply $x\succeq_r y$. When $r$ is asymmetric, $r(x,y)>0$ implies $x\succ_r y$. Hence, strict monotonicity and asymmetry jointly yield a strict form of $>$-consistency: $x> z$ and $z\succeq_r y$ imply $x\succ_r y$. This is a key observation related to our characterizations of WGARP.\footnote{When $r$ is reflexive, $>$-consistency also ensures the usual monotonicity property: $x> z$
implies $x\succ_r z$.}

The strict $>$-consistency property is a minimal form of transitivity. If $x\succ_r z$ and $z\succeq_r y$, the consumer may not be able to conclude $x\succ_r y$, perhaps, because she cannot rank these bundles with a high degree of confidence. However, if $x > z$ she should realize that $x$ is clearly superior to $z$ and deduce that $x$ is strictly preferable to any other bundle $y$ that she considers already worse than $z$. This is the normative meaning of strict $>$-consistency. Closely related consistency properties for ordered pairs have been extensively studied in recent years (see, e.g., \citet{Giar-Gre2013}, \citet{Cer-Ok2018}, \citet{nishimura2018transitive}, \citet{Cerreia-Vioglio_etal2020}, \citet{evren2021extension}).

A skew-symmetric $r$ induces a complete preference relation $\succeq_r$. \citet{shafer_nontransitive_1974} has shown that the converse also holds: Any continuous and complete binary relation $\succeq$ on $X$ can be represented by a skew-symmetric preference function $r$. In Section \ref{Sec:Quasitransitive}, we will see that this equivalence breaks down when we add strict monotonicity to the mix. Specifically, given a strictly increasing, complete, and continuous preference function $r$, the relation $\succeq_r$ may be incompatible with all strictly increasing and skew-symmetric preference functions. Thus, skew-symmetry is a stronger property than completeness within the realm of strictly increasing preference functions.

\subsection{The Coalitional Multi-Utility Preference Function}

In this subsection, we introduce a particular class of preference
functions, which is the focus of a growing literature on multi-utility representations of preference relations. 

Let $C(X)$ denote the space of continuous functions on $X.$ We equip $C(X)$ with the compact convergence topology: a sequence $\{u_n\}$ in $C(X)$ converges to $u\in C(X)$ if $u_n$ restricted to any compact set $B\subset X$ uniformly converges to $u$, as $n\rightarrow \infty$. A compatible metric $d$ is defined as $d(u,v)=\sum_{n\in \mathbb{N}} 2^{-n}d_n(u,v)/(1+d_n(u,v))$, where $d_n(u,v)\equiv \sup_{x\in B_n}|u(x)-v(x)|$ and $B_n\equiv\{x\in X:\|x\|\leq n\}$, for every $u,v\in C(X)$ and $n \in \mathbb{N}$.

A \emph{coalition structure} refers to a nonempty collection $\Omega$ of
nonempty sets $U\subseteq C(X)$. A set $U$ in $\Omega$ is called a \emph{coalition}. Each function in a coalition represents an \emph{agent}, i.e., a self of the consumer. We say that a coalition structure $\Omega$ is \emph{compact} if it is compact with respect to the Hausdorff metric $d_{H}$ induced by $d$, and if every coalition $U$ in $\Omega$ is a compact subset of $C(X)$ with respect to $d$. 

A compact coalition structure $\Omega$
induces a preference function $r^{\Omega}:X\times X\mapsto\mathbb{R}$
by the rule%
\[
r^{\Omega}(x,y)=\max_{U\in\Omega}\min_{u\in U}\left(  u(x)-u(y)\right)
\text{.}%
\]

\emph{A coalitional multi-utility (CMU) function} refers to a preference function that equals $r^{\Omega}$ for a compact coalition structure $\Omega$. We will denote the associated preference and strict preference relations by $\succeq_\Omega$ and $\succ_\Omega$, respectively.
The CMU model has been introduced by \citet{nishimura2016utility}, together with a dual $\min\max$ definition. \citet{nishimura2016utility} show that CMU is an extremely flexible multiple-selves model, making it attractive for our search of a rationalization for WGARP. Fundamentally, the CMU model does not imply the completeness or transitivity restrictions of the standard theory, which corresponds to a very special case that involves only a single coalition $\{u\}$ for a function $u \in C(X).$

By definition, we have $r^\Omega(x,y)\geq 0$ if and only if $\Omega$ contains a coalition $U$ such that $u(x)\geq u(y)$ for all $u\in U$. That is, $x$ is preferred to $y$ if the agents in at least one coalition $U$ unanimously agree that $x$ is at least as good as $y$. The next example illustrates the workings of the CMU model.

\begin{example}
\label{EX:moods}
Let $L=3$, with $x_{1}$,
$x_{2}$, and $x_{3}$ representing the consumption of vegetables, chocolate,
and meat, respectively. Consider a coalition structure $\Omega=\{U_{s},U_{h},U_{f}\}$,
where $s$, $h$ and $f$ stand for the stoic, hedonistic, and flexible coalitions (moods), respectively. 
The coalition $U_s$ consists of functions $u_{s,\alpha}(x)=\alpha x_1+0.5 x_2+(1-\alpha) x_3$ for $\alpha\in\{0.5,0.6,0.7\}$. Similarly, $U_h$ consists of $u_{h,\beta}(x)=0.5 x_1+\beta x_2+(1-\beta) x_3$ for $\beta\in\{0.5,0.6,0.7\}$, while $U_f=\{u_{f,1},u_{f,2}\}$, $u_{f,1}=0.5 x_1+0.6 x_2+0.4 x_3$, and $u_{f,2}=0.6 x_1+0.5 x_2 + 0.4 x_3$. Direct calculations show that the CMU function $r^\Omega$ is neither transitive nor complete. For example, with $x^s=(21,0,0)$, $x^h=(0,10.8,10.8)$, and $x^f=(10,10,0)$, we have
\[
\begin{array}
[c]{l}%
r^\Omega(x^{s},x^{f})\geq\min_{u\in U_{s}}\left(u(x^{s})-u(x^{f})\right)=0.5,\\
r^\Omega(x^{f},x^{h})\geq\min_{u\in U_{f}}\left(u(x^{f})-u(x^{h})\right)=0.2,\\
r^\Omega(x^{h},x^{s})\geq\min_{u\in U_{h}}\left(u(x^{h})-u(x^{s})\right)=0.3.
\end{array}
\]
In Section \ref{Sec: WGARP-Compatible CMU}, we will see that a certain property  of the coalition structure --satisfied here-- $\Omega$ guarantees asymmetry of $r^\Omega$. But the inequalities above are sufficient to conclude $x^{s}\succ_{\Omega} x^{f}\succ_{\Omega} x^{h}\succ_{\Omega} x^{s}$. Such cycles occur because whether $x\succeq_\Omega y$ is determined based on the most advantageous mood according to $x,$ while within every mood the $\min$-criterion is applied, in a conservative fashion. The $\min$-criterion, in turn, makes the relation $\succeq_\Omega$ incomplete. For example, with $y=(14,0,7)$, we have neither $y\succeq_\Omega x^f$ nor $x^f\succeq_\Omega y.$

\end{example}

Figure \ref{fig:CMUfig} below provides a chart of coalitions involved in Example \ref{EX:moods}, and their place in the $\max\min$ procedure.\footnote{A further example in Section \ref{Sec: WGARP-Compatible CMU} connects our $\max\min$ approach to voting rules routinely used in social choice problems.}

\begin{figure}[ht]
    \centering
    \begin{tikzpicture}[
     clear/.style={
         draw=none,
         fill=none
     },
     net/.style={
         matrix of nodes,
         nodes={ draw, circle, inner sep=3pt, thick },
         nodes in empty cells,
         column sep=2cm,
         row sep=0.5cm,
         font=\scriptsize,
         scale=0.75,
         every node/.style={scale=0.75}
     },
     dashededge/.style={->,>={Stealth[round]}, thick, dashed} % Style for dashed edges
    ]

    \matrix[net] (mat)
    {
    |[clear]| \parbox{0.3cm}{\centering Utilities}
        & |[clear]| \parbox{0.3cm}{\centering Coalitions}
        & |[clear]| \parbox{0.3cm}{\centering Output} \\
   
    $u_{h,0.6}(x)-u_{h,0.6}(y)$          & |[clear]|        & |[clear]| \\
    |[clear]|         & |[clear]|        & |[clear]| \\
    $u_{h,0.7}(x)-u_{h,0.7}(y)$          & $\min_{u\in U_{h}}$      & |[clear]| \\
    |[clear]|         & |[clear]|        & |[clear]| \\
    $u_{h,0.5}(x)-u_{h,0.5}(y)$          & |[clear]|           & |[clear]| \\
    |[clear]|         & $\min_{u\in U_{s}}$         & $\max_{U\in \Omega}$ \\
    $u_{s,0.6}(x)-u_{s,0.6}(y)$          & |[clear]|        & |[clear]| \\
    |[clear]|         & $\min_{u\in U_{f}}$       & |[clear]|  \\
    $u_{s,0.7}(x)-u_{s,0.7}(y)$          & |[clear]|        & |[clear]| \\
    |[clear]|         & |[clear]|        & |[clear]| \\
    };

    % Dashed lines to Uh
    \foreach \ai/\aii in {2/4, 4/4, 6/4} {
        \draw[dashededge] (mat-\ai-1) -- (mat-\aii-2);
    }

    % Dashed lines to Us
    \foreach \ai/\aii in {6/7, 8/7, 10/7} {
        \draw[dashededge] (mat-\ai-1) -- (mat-\aii-2);
    }

    % Dashed lines to Uf
    \foreach \ai/\aii in {2/9, 8/9} {
        \draw[dashededge] (mat-\ai-1) -- (mat-\aii-2);
    }

    % Dashed lines from coalitions to max
    \foreach \ai in {4,7,9} {
        \draw[dashededge] (mat-\ai-2) -- (mat-7-3);
    }
    \end{tikzpicture}
    \caption{The coalitions in Example~\ref{EX:moods}.}
    \label{fig:CMUfig}
\end{figure}

The following lemma establishes basic properties of CMU functions. In particular, we see that strict monotonicity of functions in a coalition structure is a sufficient condition for strict monotonicity of the corresponding CMU function.

\begin{lemma}
\label{Lem:Basic CMU}
 Let $\Omega$ be a compact coalition structure.
\begin{enumerate}
\item The CMU function $r^\Omega$ is continuous on $X\times X.$
\item If every $U\in \Omega$ consists of strictly increasing functions, then $r^{\Omega}(x,y)$ is strictly increasing in $x$ and
strictly decreasing in $y.$
\end{enumerate}
\end{lemma}

Toward our characterizations of WGARP, in the next section we will introduce a further property that guarantees asymmetry of a CMU function.

\section{Characterizations of WGARP \label{sec:main}}
\label{SecWGARP}

This section presents our main result on WGARP and its rationalizability by preference functions.

A preference function $r$ induces a demand correspondence defined as 
\[x_r(p,w)=\{x\in B(p,w):r(x,y)\geq 0, \; \forall y \in B(p,w)\},\]
for every price-wealth pair $(p,w)\in P\times W$. Equivalently, $x_r(p,w)$ consists of bundles that \emph{maximize} the preference relation  $\succeq_r$ on the budget set $B(p,w)$. In parallel, our rationalization notion demands consistency with maximization of a preference relation that is possibly intransitive or incomplete:

\begin{definition}
  (Preference function rationalization)
\label{PrefRat}
 A preference function $r:X\times X\mapsto\mathbb{R}$ rationalizes a dataset $O^{T}=\{(p^{t},x^{t})\}_{t\in\mathbb{T}}$ if $x^{t}\in x_r(p^{t},w^{t})$ for every $t\in\mathbb{T}$.

%previous version here 
%Consider a finite dataset $O^{T}=\{p^{t},x^{t}\}_{t\in\mathbb{T}}$
%and a preference function $r:X\times X\mapsto\mathbb{R}$. For all
%$y\in X$ and all $t\in\mathbb{T}$\ such that $p^{t}y\leq p^{t}x^{t}$,
%the dataset $O^{T}$ is rationalized by $r$ if $r(x^{t},y)\geq0$.
\end{definition}

Unlike its classical counterpart in Definition \ref{UtilityRat}, the rationalization notion above does not impose any restrictions on the structure of the preference function $r$. While this allows us to dispense with GARP, it also produces a notable difficulty: If the rationalizing preference function $r$ is intransitive or incomplete, $x_r(p,w)$ may occasionally be empty, that is, there may be no preference maximizing bundle at some price-wealth pairs. We address this issue in Sections \ref{sec:negative} and \ref{Sec:Nakamura}.\footnote{For the case of an incomplete preference function $r$, the maximization condition that we adopt in this paper agrees with \citet{gerasimou2018indecisiveness}'s approach, who models choice deferral using a choice correspondence that takes empty values occasionally. Indeed, in \citet{gerasimou2018indecisiveness}'s maximally dominant choice model, deferral occurs precisely when there is no alternative that maximizes the consumer's preference relation.  In Appendix \ref{app:Incomplete}, we also discuss an alternative specification of a demand correspondence based on \emph{undominated} alternatives, as opposed to preference maximizers.} 

A rationalizable dataset in the general sense above may well violate WGARP. Strict monotonicity and asymmetry of the rationalizing preference function turn out to be key properties that ensure compatibility with WGARP.

\begin{lemma}
\label{Lem:Asymmetry}
If a dataset $O^T$ is rationalizable by a strictly increasing and asymmetric preference function $r$, then $O^T$ satisfies WGARP.
\end{lemma}

As we noted in Section \ref{subsec:preffunctions}, a strictly increasing and asymmetric preference function $r$ is strictly consistent with the relation $>$, that is, we have $x\succ_r y$ whenever $x> z$ and $z\succeq_r y$. This property also implies than an optimal bundle $x\in x_r(p,w)$ satisfies $x\succ_r y$ for any other bundle $y$ such that $py<w$ (see Claim \ref{budget interior} in Appendix \ref{app:proofs}). If such a preference function $r$ rationalizes the dataset, we can conclude that $x^s \succ_r x^t$ whenever $p^s x^s > p^s x^t$. But then the bundle $x^s$, being strictly preferred to $x^t$, cannot belong to the budget set $B(p^t,w^t)$ where $x^t$ must be the most preferred alternative, by rationalization. This is how we deduce WGARP from rationalization by strictly increasing and asymmetric preference functions.  

Since skew-symmetry implies asymmetry, skew-symmetric and strictly increasing preference functions are also compatible with WGARP. Next, we introduce special classes of CMU functions that are compatible with WGARP, thanks to their asymmetry/skewsymmetry and strict monotonicity.

\subsection{WGARP-Compatible CMU Functions}
\label{Sec: WGARP-Compatible CMU}
 We derive asymmetry of a CMU function from the following property: 
 
 \begin{definition}
  \label{def:coherent}
   We say that a coalition structure $\Omega$ (or the associated CMU function $r^\Omega$) is \emph{coherent} if $ U\cap V \neq \emptyset$  
for any pair of coalitions $U,V\in\Omega$.
\end{definition}

Intuitively, coherence demands at least one common agent in every pair of coalitions. Such social agreement conditions have a long history in the literature on coalitional voting games. We explore these connections in Example \ref{exp:supermajority} below, and in Section \ref{Sec:Nakamura}. More recently, in the multi-utility literature, \citet{hara2019coalitional} have shown that complete preference relations can be represented by coherent coalition structures, but in a dual $\min\max$ form. By contrast, in our $\max\min$ approach, coherence is associated with asymmetry:

\begin{lemma} (Coherence)
 \label{lem:coherent}
  A coherent CMU function $r^\Omega$ is asymmetric.
\end{lemma}

If all agents $u$ in a coalition $U$ agree that $x$ is strictly better than $y$, then coherence guarantees that any other coalition $V$ also contains such an agent, implying $\min_{v\in V} \left(v(y)-v(x)\right)<0$. This observation underlies the connection between coherence and asymmetry properties in our $\max\min$ approach.

 Lemma \ref{lem:coherent} clarifies why the CMU function in Example \ref{EX:moods} is asymmetric:\bigskip
 
\noindent \textbf{Example \ref{EX:moods} (Continued).} {\em The coalition structure $\Omega=\{U_{s},U_{h},U_{f}\}$ is coherent. Indeed, by definitions, $u_{s,0.5}=u_{h,0.5}$, $u_{h,0.6}=u_{f,1}$, and $u_{f,2}=u_{s,0.6}$. Consequently, we have
\[
u_{s,0.5}\in U_s\cap U_h,\quad u_{h,0.6}\in U_h\cap U_f \quad\text{and}\quad u_{f,2} \in U_f\cap U_s.
\]
}

A consumer with multiple selves is effectively facing a social choice problem that involves multiple agents with conflicting interests. Simple and super-majority voting rules adapted to such a consumer provide an interesting class of coherent CMU functions:

\begin{example}
\label{exp:supermajority}
Pick $n$ strictly increasing functions $u_1,\ldots,u_n$ in $C(X)$. For any set of indices $J$ in $N\equiv \{1,\ldots,n\}$, let $U_J=\{u_i:i\in J\}$. Given a fixed integer $m>n/2,$ the collection of all coalitions with cardinality $m$ or greater is the set $\Omega (m) = \{U_J:J\subseteq N, |J|\geq m \}.$ The coalition structure $\Omega (m)$ is coherent because among $n$ agents, any pair of coalitions with greater than $n/2$ members must have a common member. With this specification, we have $r^{\Omega(m)}(x,y)\geq 0$ if and only if there exist a set $J\subseteq N$ such that $|J|\geq m$ and $u_i(x)\geq u_i(y)$ for every $i\in J.$ This is a majority rule with the threshold level $m$. 
\end{example}

As noted earlier, the CMU function in Example \ref{EX:moods} is incomplete. Typically, the CMU function in Example \ref{exp:supermajority} is also incomplete whenever the majority threshold $m$ is greater than $(1+n)/2$. The special case $m=(1+n)/2$ for an odd number $n$ corresponds to the simple-majority rule, which gives a complete and coherent CMU function. 

The examples above show that coherence is logically independent from completeness and skew-symmetry. Next, we construct a special class of coherent CMU functions that also prove to be skew-symmetric. This will allow us to conclude that in coherent CMU rationalizations, assuming completeness does not cause a loss of generality.

Given a natural number $n$, consider an $n\times n$ matrix $\mathcal{M}=[u_{ij}]$ of functions $u_{ij}\in C(X)$ for $i,j\in N=\{1,\ldots,n\}$. Assume further that $\mathcal{M}$ is \emph{symmetric}, that is $u_{ij}=u_{ji}$ for every $i,j\in N.$ 

%An ordered pair of bundles $(x,y)$ is associated with a utility difference matrix $\mathcal{M}(x,y)=[u_{ij}(x)-u_{ij}(y)]\in \mathbb{R}^{n\times n}.$

Let us denote by $\lambda$ and $\mu$ generic elements of the $n-1$ dimensional simplex $\Delta=\{\lambda\in\mathbb{R}_{+}^{n}:\sum_{i=1}^{n}\lambda_{i}=1\}$. For every $\lambda,\mu\in\Delta,$ define a function $u_{\lambda,\mu}= \sum_{i\in N}\sum_{j\in N}\lambda_{i}\mu_{j}u_{ij}$. Consider a coalition structure $\Omega(\mathcal{M})$ defined as 
\begin{equation}
\label{def:Omega(M)}
\Omega(\mathcal{M}) = \{U_\lambda:\lambda\in \Delta\},\quad \text{where}\quad U_\lambda= \{u_{\lambda,\mu}:\mu \in \Delta\},
\end{equation}
for every $\lambda \in \Delta.$

\begin{lemma}
\label{lem:CMU-skewsymmetric}
For a symmetric matrix $\mathcal{M}=[u_{ij}]\in C(X)^{n\times n}$, the coalition structure $\Omega(\mathcal{M})$ is compact and coherent. Moreover, the associated CMU function $r^{\Omega(\mathcal{M})}$ is skew-symmetric. 
\end{lemma}

We establish compactness of $\Omega(\mathcal{M})$ using routine arguments based on continuity of the function $(\lambda,\mu)\mapsto u_{\lambda,\mu}$ that maps the compact set $\Delta\times \Delta$ into $C(X)$.

%$U_\lambda$ are compact because they are bounded and closed subsets of a finite dimensional subspace in $C(X)$. Moreover, $\lambda\mapsto U_\lambda$ is a continuous function that maps $\Delta$ into compact subsets of $C(X)$ equipped with the Hausdorff metric. These observations imply that $\Omega(\mathcal{M})$ is a compact coalition structure.

Coherence of $\Omega(\mathcal{M})$ follows from symmetry of the matrix $\mathcal{M}=[u_{ij}]$. 
For any pair $\lambda,\hat{\lambda}\in \Delta$, we have
\begin{equation}
\label{eq:symmetry}
u_{\lambda,\hat{\lambda}} = \sum_{i\in N}\sum_{j\in N}\lambda_{i}\hat{\lambda}_{j}u_{ij}=\sum_{i\in N}\sum_{j\in N}\lambda_{i}\hat{\lambda}_{j}u_{ji}=\sum_{i\in N}\sum_{j\in N}\hat{\lambda}_{i}\lambda_{j}u_{ij}=u_{\hat{\lambda},\lambda}.
\end{equation}
Here, the second equality follows from the symmetry property $u_{ij}=u_{ji}$, and the third one from a change of indices. By equation (\ref{eq:symmetry}), the function $u_{\lambda,\hat{\lambda}}=u_{\hat{\lambda},\lambda}$ belongs to both $U_\lambda$ and $U_{\hat{\lambda}}$, proving coherence of $\Omega(\mathcal{M})$. 

Finally, we derive skew-symmetry of $r^{\Omega(\mathcal{M})}$ from the minimax theorem and the symmetry of $\mathcal{M}$, which allow us to show that $r^{\Omega(\mathcal{M})}$ can equivalently be expressed in the $\min\max$ form, that is, $r^{\Omega(\mathcal{M})}(x,y)=\min_{\lambda\in \Delta} \max_{u \in U_\lambda} (u(x)-u(y))$ for every $x,y\in X$. We present the details of this proof in Appendix \ref{app:proofs}.

\subsection{Main Result}

The next theorem shows that WGARP, as an empirically testable property, is equivalent to rationalizability of the dataset by a strictly increasing and coherent CMU function $r^\Omega$. Without loss of generality, we can also assume that the rationalizing function $r^\Omega$ is skew-symmetric, and each coalition in $\Omega$ consists of concave functions.

\begin{theorem}
 \label{thm:WGARPCharacterization}
Consider a finite dataset $O^{T}=\{(p^{t},x^{t})\}_{t\in\mathbb{T}}$. The following
statements are equivalent:
\begin{enumerate}
\item 
The dataset $O^{T}$ can be rationalized by an asymmetric, strictly increasing, and continuous preference function $r:X\times X \mapsto \mathbb{R}$.
\item The dataset $O^{T}$ can be rationalized by a coherent, strictly increasing,  and continuous CMU function.
\item The dataset $O^{T}$ can be rationalized by a skew-symmetric, coherent, and continuous CMU function $r^{\Omega}$ such that each coalition in $\Omega$ consists of strictly increasing and concave functions.
\item The dataset $O^{T}$ satisfies WGARP.
\item There exist numbers $R^{s,t}$ and $\lambda^s_{s,t},\lambda^t_{s,t}>0$ for all
$s,t\in\mathbb{T}$ with $R^{s,t}=-R^{t,s}$ and $\lambda^s_{s,t}=\lambda^s_{t,s}$
such that inequalities:
\[
R^{s,t}\geq\lambda^s_{s,t}p^{s}(x^{s}-x^{t}),
\]
hold for all $s,t\in\mathbb{T}$.
\item There exist numbers $V^{s,t}$ for all $s,t\in\mathbb{T}$ with $V^{s,t}=-V^{t,s}$\ such
that inequalities:
\begin{eqnarray*}
 &  & \text{if }p^{s}(x^{s}-x^{t})\geq0,\text{ then }V^{s,t}\geq0,\\
 &  & \text{if }p^{s}(x^{s}-x^{t})>0,\text{ then }V^{s,t}>0,
\end{eqnarray*}
hold for all $s,t\in\mathbb{T}$.
%\item 
%\label{thm:WGARPCharacterization(v)}
%The dataset $O^{T}$ can be rationalized by a coherent, skew-symmetric, strictly %increasing,  and continuous CMU preference function.
%\item 
%\label{thm:WGARPCharacterization(vi)}
%The dataset $O^{T}$ can be rationalized by a coherent, strictly increasing,  and %continuous CMU preference function.
\end{enumerate}
\end{theorem}

Statement (i) in Theorem \ref{thm:WGARPCharacterization} replaces a strictly increasing utility function in Afriat's theorem with a strictly increasing and asymmetric preference function $r$. This characterizes the difference between GARP and WGARP in terms of the structural properties of the rationalizing preference functions.\footnote{In statement (i), the strict monotonicity condition can be replaced with the following local non-satiation property: For any $\varepsilon>0$ and $x,y\in X$ with $x\sim_r y,$ there exists an $x' \in X$ such that $\|x'-x\|<\varepsilon$ and $r(x',y)>0$. For brevity, we focus on strictly monotone preference functions.  }

Theorem 1 also shows that under WGARP, continuity or completeness of the rationalizing preference functions have no testable implications, just as in Afriat's theorem. An important difference lies in statement (iii). While we can always use concave functions $u$ to construct a rationalizing CMU function $r^\Omega,$ this does not imply concavity of the preference function $r^\Omega(x,y)$ in $x$, nor does it make the preference relation $\succeq_\Omega$ convex.\footnote{Concavity of the functions $u$ in statement (iii) only implies that for each bundle $x$, the upper contour set $\succeq_\Omega(x)\equiv\{y\in X: y\succeq_\Omega x\}$ is \emph{star shaped}: For any $y$ in $\succeq_\Omega(x)$ the interval between $y$ and $x$ is also contained in $\succeq_\Omega(x)$.} In fact, in the present setup, rationalizability of a dataset by a convex preference relation requires an axiom that is stronger than WGARP. (More on this in Section \ref{sec:negative} and Appendix \ref{app:Non-Convexity}.)    

The most involved part of Theorem \ref{thm:WGARPCharacterization} is the derivation of statement (iii) from WGARP. Consider a dataset $O^{T}$ that satisfies WGARP. For any pair of indices $s,t\in \mathbb{T}$, Afriat's theorem delivers a concave and strictly increasing function $u_{st}\in C(X)$ that rationalizes the choices of $x^s$ and $x^t$ from the budget sets $B(p^s,w^s)$ and $B(p^t,w^t)$, respectively. Define a symmetric $T\times T$ matrix $\mathcal{M}=[u_{st}]$ such that $u_{st}=u_{ts}$ for every $s,t\in\mathbb{T}$. In the proof of statement (iii), we use the coalition structure $\Omega(\mathcal{M})$ induced by the matrix $\mathcal{M}$ and the $T-1$ dimensional simplex $\Delta$. From Lemma $\ref{lem:CMU-skewsymmetric}$, we know that the CMU function $r^{\Omega(\mathcal{M})}$ is skew-symmetric and coherent. Note also that the set $U_\lambda$ on the  right side of expression (\ref{def:Omega(M)}) for the degenerate weight distribution $\lambda=\bold{1}_{s}$ equals the set $U_s=\{\sum_{t\in \mathbb{T}}\mu_{t}u_{st}:\mu\in\Delta\}$. Since $x^s$ maximizes $u_{st}$ on $B(p^s,w^s)$ for every $t\in \mathbb{T},$ it also maximizes every function $u\in U_s$ on the same budget set. It thus follows that
\[
\label{r func}
r^\Omega(x^s,y)=\max_{\lambda\in\Delta}\min_{u\in U_\lambda}\left((u(x^s)-u(y)\right)\geq\min_{u\in U_s}\left((u(x^s)-u(y)\right)\geq 0,
\]
for every $y \in B(p^s,w^s)$. We thereby conclude that the CMU function $r^{\Omega(\mathcal{M})}$ rationalizes the dataset $O^T$.

While statement (iii) in Theorem \ref{thm:WGARPCharacterization} combines coherence with skew-symmetry, each of these properties are separately sufficient for compatibility of a rationalized dataset with WGARP. In statement (ii), dropping the skew-symmetry (and hence, completeness) condition allows us to rationalize the dataset with a finite coalition structure $\Omega = \{V_s:s\in \mathbb{T}\},$ where $V_s\equiv \{u_{st}:t \in \mathbb{T}\}$. 

In statements (v) and (vi) of Theorem \ref{thm:WGARPCharacterization}, the interpretations of the variables $R^{s,t},\lambda^s_{s,t}$, and $V^{s,t}$ remain the same as in Afriat's theorem: $R^{s,t}=V^{s,t}$ corresponds to the utility difference $u_{st}(x^s)-u_{st}(x^t)$, while $\lambda^s_{s,t}$ represents the marginal utility of income in observation $s$ associated with the local utility function $u_{st}$.  

\subsection{Compensated Law of Demand}
Even the most general form of rationalizing functions in Theorem \ref{thm:WGARPCharacterization} comply with the compensated law of demand. That is, the price of a good and the demand for it tend to move in opposite directions, net of the income effect induced by the price change:

\begin{lemma}
\label{lem:CompensatedLD}
Let $r$ be an asymmetric and strictly increasing preference function. Suppose a bundle $x$ belongs to $x_r(p,w)$ for some $(p,w)\in P\times W$. Pick any $p'\in P$ and set $w'\equiv p'x$. Then for any $x' \in x_r(p',w')$ we have
\begin{equation}
(p'-p)(x'-x)\leq 0.\label{eq:CLD}
\end{equation} 
\end{lemma}

The proof of Lemma \ref{lem:CompensatedLD} easily follows from Walras' law, and the fact that an asymmetric and strictly increasing preference function $r$ induces a demand correspondence that satisfies WGARP. 

In line with Example \ref{exp:CLD}, the converse also holds: If a dataset $\{(p,x),(p',x')\}$ violates WGARP, so that $px'\leq px$ and $p'x < p'x'$, then inequality (\ref{eq:CLD}) fails. 

Next, we illustrate how such violations of WGARP may occur under alternative CMU specifications.

\subsection{Justifiable CMU and Violations of WGARP}
\label{Sec:Quasitransitive}

In this section, we show that a well-known CMU function typically generates violations of WGARP, despite being continuous, complete, and strictly increasing. This highlights the special structure of the preference functions constructed in Theorem \ref{thm:WGARPCharacterization}, which feature strict monotonicity simultaneously with asymmetry/skew-symmetry. 

Consider a coalition structure $\Omega' = \{\{u\}:u\in U\}$, where $U\subseteq C(X)$ is a nonempty compact set. Here each agent with a utility function $u$ acts as a separate coalition. The associated preference function $r^{\Omega'}(x,y)\equiv \max_{u\in U} (u(x)-u(y))$ induces the following preference relation:
\begin{equation}
\label{justifiable}
x \succeq_{\Omega'} y\quad \text{if and only if}\quad \max_{u\in U} (u(x)-u(y))\geq 0,
\end{equation}
for every $x,y \in X$.
 
The history of this particular CMU model goes back to \citet{Aiz-Mal1981}'s choice-theoretic study. A more recent contribution is due to \citet{Leh-Tep2011}, who focus on multiple priors in the context of decision making under uncertainty. Following Lehrer and Teper, representations of the form (\ref{justifiable}) became known as the \emph{justifiable} preference model. Proposition 4 in \citet{nishimura2016utility} shows that any continuous, complete and quasitransitive\footnote{A binary relation $\succeq$ is said to be \emph{quasitransitive} if its strict part $\succ$ is transitive.} binary relation $\succeq$ on $X$ can be represented as in (\ref{justifiable}), possibly with a more general set of functions, $U$.

The next lemma shows that aside from degenerate cases, the justifiable CMU model necessarily violates WGARP.

\begin{lemma}
\label{lem:No-WGARP}
In representation (\ref{justifiable}), suppose every function $u$ in the set $U$ is quasiconcave, continuous, and strictly increasing on $X$. Let us write $r$ in place of the associated CMU funtion $r^{\Omega'}$. One of the following statements holds true: 
\begin{enumerate}
\item \label{lem:No-WGARP(i)} Every dataset that can be rationalized by $r$ satisfies GARP.
\item \label{lem:No-WGARP(ii)} There exist price-wealth pairs $(p^1,w^1),\;(p^2,w^2)$ in $P\times W$ and bundles $x^1,\; x^2$ such that $x^1\in x_r(p^1,w^1), \; x^2\in x_r(p^2,w^2),\; p^1x^2<w^1,$ and $p^2x^1<w^2$.
\end{enumerate}
\end{lemma}

 Thus, if a justifiable CMU model is not observationally equivalent to standard utility maximization, and if the functions in the representing set $U$ satisfy the regularity conditions in Lemma \ref{lem:No-WGARP}, then the model generates predictions that violate WGARP, as in statement (ii). 

Note that the coalition structure $\Omega' = \{\{u\}:u\in U\}$ is \emph{not} coherent unless $U$ consists of a single function. In the latter degenerate case, the implied behavior is fully rationalizable, as in statement (i). The same conclusion holds if all functions in $U$ produce the same demand correspondence. In the proof of Lemma \ref{lem:No-WGARP}, we show that violations of WGARP will necessarily occur in all other cases.

It is also important to note that given a compact set $U$ that satisfies the conditions in Lemma \ref{lem:No-WGARP}, the associated justifiable CMU function $r^{\Omega'}$ is strictly increasing, complete, and continuous. Yet, by Lemmas \ref{Lem:Asymmetry} and \ref{lem:No-WGARP}, we see that typically the preference relation $\succeq_{\Omega'}$ is not representable by a skew-symmetric and strictly increasing preference function. It follows that within the class of strictly increasing preference functions, skew-symmetry is a stronger property than completeness, and only the former implies WGARP.

\section{Discussion: The Possibility of Empty Demand \label{sec:negative}}
\label{Sec:negative exp}

We now turn our attention to some technical difficulties. Given the aforementioned association between the justifiable CMU model and quasitransitive preference relations, Lemma \ref{lem:No-WGARP} can be interpreted as saying that a typical preference function $r$ that complies with WGARP is either incomplete, or violates quasitransitivity, i.e., $\succ_r$ contains strict preference cycles. Naturally, both incompleteness and strict preference cycles may lead to a demand correspondence that takes empty values on some price-wealth pairs.

As an important exception, a complete preference function that is quasiconcave (in the first $L$ variables) is known to induce a nonempty-valued demand correspondence (see \citet{shafer_nontransitive_1974},
and \citet{Sonnenschein1971}). However, WGARP is not sufficient to guarantee the existence of a rationalizing quasiconcave preference function that is also asymmetric and strictly increasing. Indeed, in Appendix \ref{app:Non-Convexity} we show that a strictly increasing and asymmetric preference function that rationalizes the dataset in Example \ref{ex:WGARPnotGARP} cannot be quasiconcave.\footnote{The example serves to refute two long-standing conjectures: the one in  \citet{kihlstrom_demand_1976-1} asserting that WGARP would imply convex preferences, and the Samuelson ``eternal darkness'' conjecture, that convexity of preferences is not testable in finite datasets (see \citet{Pol-Ren2016} and references therein.)} From the same example, it also follows that it might be impossible to make out-of-sample demand predictions without violating WGARP, even if the original dataset complies with WGARP:

\begin{example}\label{ex:DemandCounterfactuals}
Let $O^{3}$ denote the dataset in Example \ref{ex:WGARPnotGARP}, with prices $p^{1}=(4,1,5)$, $p^{2}=(5,4,1)$,
$p^{3}=(1,5,4)$, and bundles $x^{1}=(4,1,1)$, $x^{2}=(1,4,1)$,
$x^{3}=(1,1,4)$. As noted in Example \ref{ex:WGARPnotGARP}, the dataset $O^{3}$ satisfies
WGARP.

Define a price-wealth pair $(\hat{p},\hat{w})$ as $\hat{p}=\frac{1}{3}(p^{1}+p^{2}+p^{3})$ and $\hat{w}=\frac{1}{3}(w^{1}+w^{2}+w^{3})$. Then $B(\hat{p},\hat{w})\subseteq \bigcup_{t=1}^{3}B(p^t,w^t)$, that is, for any bundle $x$ with $\hat{p}x\leq\hat{w}$, there exists a $t\in \{1,2,3\}$ such that $p^tx\leq w^t$.

Pick any asymmetric and strictly increasing preference function $r$ that rationalizes $O^{3}$. We claim that $x_r(\hat{p},\hat{w})=\emptyset$. Otherwise, we can select a point $x^4\in x_r(\hat{p},\hat{w})$ and form a larger dataset $O^4\equiv O^3\cup\{(p^4,x^4)\}$ where $p^4\equiv \hat{p}$. Then, by construction, $r$ rationalizes the dataset $O^4$. As $r$ is asymmetric and strictly increasing, this also implies that $O^4$ satisfies WGARP. Note, however, that $p^4=\hat{p}=\frac{1}{3}(10,10,10)$, $w^4=\hat{w}=22$, and $p^4x^t<w^4$ for every $t=1,2,3$. This shows that $O^4$ violates WGARP because $x^4$ belongs to at least one of the original budget sets $B(p^t,w^t)$ for some $t\in \{1,2,3\}$.   

\end{example}

How can we strenghthen WGARP to solve such existence problems? In view of the remarks above, a possible direction is to seek an axiom that guarantees rationalization by a quasiconcave preference function that is also asymmetric and strictly increasing.  In Appendix \ref{app:Non-Convexity}, we discuss this sort of an axiom due to \citet{john_concave_2001}, which actually implies a stronger form of rationalization by a
concave preference function. It remains an open question to determine whether John's axiom can be weakened if one demands a more general quasiconcave function. We also leave it for future research to study what structural assumptions on CMU functions are needed to generate convex preference relations.\footnote{Theorem 1.a of \cite{nishimura2016utility} shows that CMU functions in a general form are able to represent any reflexive binary relation. Yet, little is known about the structure of quasiconcave CMU functions.} 

Next, we briefly explore a different path that provides alternative existence results.

\section{Nonempty Demand: A Stronger Form of Acyclicity \label{sec:kacyclic}}
\label{Sec:Nakamura}
As we have seen in Section \ref{SecWGARP}, CMU representations provided by Theorem \ref{thm:WGARPCharacterization} subsume simple and super majority voting rules adapted to a consumer with multiple selves. By now, there is a large literature on core-existence results in the context of voting games (a recent survey is provided by \cite{martin2013social}). In this section, we illustrate how one can exploit the findings of this literature in order to establish the existence of preference-maximizing bundles under CMU representations. More specifically, we will use a core-existence result of \citet{schofield1984social}, based on an extension of the coherence notion, in order to establish the possibility of counterfactual demand predictions that comply with WGARP. In view of Example \ref{ex:DemandCounterfactuals}, this necessitates an axiom that is stronger than WGARP. We focus on the following acyclicity property:

\begin{axiom} ($k$-acyclicity)
For a fixed integer $k\geq 2$, there is no chain $(x^{t_{1}},x^{t_{2}},\cdots,x^{t_{k}})$ of elements of $X$ such that $x^{t_{1}}\succeq^{R,D}x^{t_{2}}\succeq^{R,D}\cdots\succeq^{R,D}x^{t_{k}}$ and $x^{t_{k}}\succ^{R,D}x^{t_{1}}$.
\end{axiom}

WGARP is a special form of the $k$-acyclicity axiom with $k=2.$ Larger choices of $k$ rule out longer cycles, and enhance the level of consistency in the dataset. However, unlike GARP, the $k$-acyclicity axiom does not rule out cycles of arbitrary length.\footnote{The $k$-acyclicity axiom can be implemented in polynomial time, using a simplified version of Warshall's algorithm. See \cite{varian_nonparametric_1982} for a discussion of Warshall's algorithm in the context of revealed preferences.} 

In Theorem \ref{thm:WGARPCharacterization}, we have seen that WGARP is equivalent to rationalizability by a coherent CMU function. How can we extend the coherence notion in order to characterize the $k$-acyclicity axiom, in general? Below, we provide an answer based on the notion of a Nakamura number, which is a measure of social agreement or collegiality that has been related to the core of voting games (see \cite{schofield1984social}, and \cite{martin2013social}). 

In our setup, given a finite\footnote{We say that a CMU function $r^\Omega$ is \emph{finite} if $\Omega$ consists of finitely-many coalitions each containing finitely-many functions.} CMU function $r^\Omega$, the \emph{Nakamura number} $\nu(\Omega)$ is defined as follows:
\[
\nu(\Omega)=\min\{|\hat{\Omega}| : \hat{\Omega} \subseteq \Omega \enspace \text{and} \enspace \cap_{U\in \hat{\Omega}}U = \emptyset\}.
\]
Put differently, $\nu(\Omega)$ is the minimum integer $k'$ such that any subcollection $\hat{\Omega}$ with fewer than $k'$ coalitions feature a common agent shared by all coalitions in $\hat{\Omega}$. Thus, for a coherent CMU preference function $r^\Omega$, we have  $\nu(\Omega)\geq 3$. Also, by convention, $\nu(\Omega)=\infty$ if all coalitions in $\Omega$ share a common function. 

The next theorem provides a characterization of the $k$-acyclicity axiom:
\begin{theorem}
\label{thm:Acyclicityk}
Consider a finite dataset $O^{T}=\{(p^{t},x^{t})\}_{t\in\mathbb{T}}$ and any integer $k\geq 2$. The following
statements are equivalent:
\begin{enumerate}

    \item The dataset $O^{T}$ can be rationalized by a finite CMU function $r^\Omega$ such that $\nu(\Omega)\geq k+1$, and every coalition $U\in \Omega$ consists of strictly increasing functions.

    \item The dataset $O^{T}$ satisfies $k$-acyclicity.

\end{enumerate}
\end{theorem}
This result shows that $k$-acyclicity exhausts the empirical content of a coherent CMU function with Nakamura number $\nu\geq k+1$. The related equivalence in Theorem~\ref{thm:WGARPCharacterization} becomes a particular case with $k=2$.

A voting game with $L$ dimensional vector outcomes is known to have a nonempty core provided that the collection of permissible coalitions has a Nakamura number that exceeds $L$, and the set of feasible outcomes is compact and convex (see \citet{schofield1984social}). This leads to the following possibility result on out-of-sample predictions (nonempty demand):

\begin{theorem}
\label{thm:Nak} 

Suppose that a finite dataset $O^{T}=\{(p^{t},x^{t})\}_{t\in \mathbb{T}}$ satisfies $L$-acyclicity. Then
for any $(p,w)\in P\times W$, there
exists a bundle $x^{*}\in X$ such that $px^{*}=w$, and $O^{T+1}\equiv O^{T}%
\cup\{(p,x^{*})\}$ satisfies WGARP.
\end{theorem}

A CMU function $r^\Omega$ with a large Nakamura number is typically incomplete, as in a supermajority voting rule. Such voting rules may also exhibit cycles, but \citet{schofield1984social}'s theorem guarantees the existence of a core outcome. Given an out-of-sample budget set $B(p,w)$, the corresponding core outcome is a bundle $x^*$ such that $r^\Omega(y,x^*)>0$ does not hold for any $y \in B(p,w)$, where the CMU function $r^\Omega$ is obtained from Theorem~\ref{thm:Acyclicityk}. This is a maximality criterion that is distinct from our preference maximization condition. Nevertheless, Theorem \ref{thm:Nak} shows that the maximal/core prediction $x^*$ combined with the bundles observed in the original dataset $O^{T}$ become WGARP-compatible.  

To summarize, if we replace WGARP with $L$-acyclicity, it becomes possible to make out-of-sample demand predictions that are compatible with WGARP. In this case, the extended
dataset $O^{T +1}$ also becomes rationalizable as in Theorem \ref{thm:WGARPCharacterization}. To this end, we have used \citeauthor{schofield1984social}'s (\citeyear{schofield1984social}) theorem. In principle, the existence problems discussed in Section \ref{sec:negative} may have further solutions based on the voting literature.\footnote{We are very grateful to Chris Chambers for pointing out the connection to the literature on voting games.} 

It should also be noted that $L$-acyclicity is less demanding than GARP. The following example illustrates this point.\footnote{\citet{Kamiya1963} uses essentially the same example to make a similar
point about SARP. Our example differs only by one data point and yields a GARP violation.}

\begin{example} Let $L=3$, and consider a dataset $O^{4}$ with
prices\ $p^{1}=(1,1,0.5),$ $p^{2}=(1,1,1),$ $p^{3}=(1,0.5,1),$ $p^{4}%
=(1,2.1,2),$ and bundles $x^{1}=(8,1,8),$ $x^{2}=(5,5,6),$ $x^{3}=(5,6,5),$
$x^{4}=(8,8,1)$. Let us collect the observed prices as row vectors in a
$4\times3$ matrix $\hat{P}$ and the bundles as column vectors in a $3\times4$ matrix
$\hat{X}$:
\[
\hat{P}=\left[
\begin{array}
[c]{l}%
p^{1}\\
p^{2}\\
p^{3}\\
p^{4}%
\end{array}
\right]  =\left[
\begin{array}
[c]{ccc}%
1 & 1 & 0.5\\
1 & 1 & 1\\
1 & 0.5 & 1\\
1 & 2.1 & 2
\end{array}
\right]  ,\quad\hat{X}=\left[  x^{1},x^{2},x^{3},x^{4}\right]  =\left[
\begin{array}
[c]{cccc}%
8 & 5 & 5 & 8\\
1 & 5 & 6 & 8\\
8 & 6 & 5 & 1
\end{array}
\right]  .
\]

Then $E\equiv \hat{P}\hat{X}=\left[p^{s}x^{t}\right]$ is a $4\times4$ matrix, calculated as%
\[
E=\left[
\begin{array}
[c]{cccc}%
\mathbf{13} & 13 & 13.5 & 16.5\\
17 & \mathbf{16} & 16 & 17\\
16.5 & 13.5 & \mathbf{13} & 13\\
26.1 & 27.5 & 27.6 & \mathbf{26.8}%
\end{array}
\right]  .
\]
The bold entry in the $s$-th row of the matrix $E$ is the actual expenditure,
$p^{s}x^{s}$, in the observation $s$. More generally, the $t$-th entry in row
$s$ is the cost of the bundle $x^{t}$ at the price vector $p^{s}$. Thus, each
row of $E$ yields a unique direct revealed preference, as follows:%
\[
x^{1}\succeq^{R,D}x^{2},\quad x^{2}\succeq^{R,D}x^{3},\quad x^{3}\succeq
^{R,D}x^{4},\quad\text{and}\quad x^{4}\succ^{R,D}x^{1}.
\]
It follows that $O^{4}$ violates GARP. Yet, it is easy to check that any
subsample of size $3$ satisfies GARP, which means that $O^{4}$ satisfies
$3$-acyclicity. 

%In this example, Theorem \ref{thm:Nak} allows us to make a demand
%prediction for any out-of-sample budget set. That is, coming from the $4$-point dataset consisting of observations $(p^t,x^t)$ $t=1,2,3,4$, one can take any additional budget set $B(p,w)$ that may be of interest, find a bundle $x^*$, and a WGARP-consistent preference function $r$ that rationalizes the choice of $x^*$ from $B(p,w)$ simultaneously as the original $4$ observations. The preference function $r$ is delivered by Theorem \ref{thm:Acyclicityk} and does not depend on the choice of the budget set $B(p,w)$, but it may influence the demand prediction $x^{*}$. Needless to say, we could not use a standard utility function for the same purpose because the initial observations violate GARP.
\end{example}

\section{On Preference Maximization and Dual Problems}
\label{sec:textbook}
In this section, we first study analytical features of the consumer's problem with a general preference function. For the case of a CMU function, we relate this preference maximization problem to an alternative optimization exercise concerned with core/unblocked bundles. Then we study an expenditure minimization problem and establish its equivalence to the primary problem of preference maximization.

\subsection{Computation of Solutions}
\label{sec:Pref-Max}
Fix an asymmetric, strictly increasing, reflexive, and continuous preference function $r$. Given a budget set $B(p,w)$, the demand correspondence $x_r(p,w)$ consists of preference-maximizing bundles $x^*\in B(p,w)$ such that $r(x^*,y)\geq 0$ for every $y\in B(p,w)$. 

For each $\bar{x}\in X$, let $\Phi(\bar{x})$ denote the set of bundles that solve the following problem: 
\[
\min r(\bar{x},y) \quad\text{s.t.} \quad y\in B(p,w).    \tag{P}\label{Prb:P}
\]

Asymmetry and reflexivity of $r$ imply $r(x,x)=0$ for every $x\in X$. Thus, a bundle $x^*$ in $B(p,w)$ satisfies the preference maximization condition $x^*\in x_r(p,w)$ if and only if $y=x^*$ solves the problem (\ref{Prb:P}) at $\bar{x}=x^*$.\footnote{When $r$ is differentiable in its last $L$ coordinates, with the derivative $\nabla_2r(x,y)$, necessary first-order conditions for preference maximization at a point $x^*\gg 0$ take a familiar form: $-\nabla_2r(x^*,x^*)=\lambda p$ for some $\lambda \geq 0$. This equation also becomes sufficient for optimality of a bundle $x^*$ when $r(x^*,\cdot)$ is quasiconvex and $\lambda >0$.} Put differently, we have
\[
x_r(p,w)=\{x^*\in B(p,w): x^*\in \Phi(x^*)\}.
\]
This set is nonempty if and only if the minimum value function $\phi(\bar{x})$ in problem (\ref{Prb:P}) reaches the value $0$ at some $\bar{x}\in B(p,w)$. That is, $x_r(p,w)=\{\bar{x}\in B(p,w):\phi(\bar{x})=0\}$.

When the function $r(x,y)$ is convex in $y$, one can solve the problem (\ref{Prb:P}) with the usual methods, for each fixed bundle $\bar{x}\in B(p,w)$, and then search for a fixed point of the $\arg\min$ correspondence $\Phi(\bar{x})$. Alternatively, one can find the maximizers of the value function $\phi(\bar{x})$ on the budget set $B(p,w)$.

Let us now focus on a coherent CMU function $r=r^\Omega$, where each coalition $U$ consists of concave, strictly increasing, and continuous functions $u$ on $X$. For every $U\in \Omega,$ define another function $r_U:X\times X\mapsto \mathbb{R}$ by the rule $r_U(x,y)=\min_{u\in U} (u(x)-u(y))$. The function $r_U(x,y)$ is concave in $x$ because of the concavity of the component functions in a coalition $U$. However, in general, $r_U(x,y)$ is neither concave nor convex in $y$.

By definition, the CMU function $r^\Omega$ satisfies $r^\Omega(x,y)=\max_{U\in\Omega} r_U(x,y)$ for every $x,y\in X$. The corresponding minimization problem (\ref{Prb:P}) has been extensively studied in the optimization literature,\footnote{See \citet{tsevendorj2001piecewise}, \citet{gaudioso2018minimizing}, and references therein. More precisely, this literature studies the minimization of piecewise concave functions on compact, convex sets. A real-valued function on a convex set is said to be \emph{piecewise concave} if it can be expressed as the maximum of finitely-many concave functions that are not necessarily differentiable.} under the additional assumption that each $r_U(x,y)$ is concave in $y$, which practically requires affine component functions $u$.  Even in this relatively well-behaved case, $r^\Omega(x,y)$ is not convex in $y$, and consequently, sufficient first-order conditions become more complicated than the case of a standard minimization problem with a convex objective function (see \citet{tsevendorj2001piecewise}).

Fortunately, a dual problem is more tractable:
\[
\max r^\Omega(y,\bar{x}) \quad\text{s.t.} \quad y\in B(p,w).    \tag{C}\label{Prb:C}
\]
By asymmetry, $r^\Omega(x^*,y)\geq 0$ implies $r^\Omega(y,x^*)\leq 0$. Thus, given a preference maximizer $x^*\in x_r(p,w)$, the point $y=x^*$ also solves the problem (\ref{Prb:C}) at $\bar{x}=x^*$. Moreover, since $r^\Omega(y,x^*)=\max_{U\in\Omega} r_U(y,x^*)$, we have
\[
r^\Omega(y,x^*)\leq 0\quad \Leftrightarrow \quad r_U(y,x^*)\leq 0\; \forall U\in \Omega.
\]

If a bundle $x^*$ solves problem (\ref{Prb:C}) at $\bar{x}=x^*,$ there does not exist another bundle $y\in B(p,w)$ such that $r^\Omega(y,x^*)>0$. In other words, there do not exist a coalition $U\in \Omega$ and a bundle $y\in B(p,w)$ such that $r_U(y,x^*)>0$. This makes $x^*$ a \emph{core} alternative that is not ``blocked'' by any coalition, as in our earlier discussion of Theorem \ref{thm:Nak}. In general, this core property is less demanding than preference maximization, but problems (\ref{Prb:P}) and (\ref{Prb:C}) become equivalent in the case of a skew-symmetric preference function $r$.

Statements (i)-(iii) in the following lemma summarize key properties of problem (\ref{Prb:C}).
\begin{lemma}
\label{lem:pref_max}
Consider a coherent CMU function $r=r^\Omega$, and suppose that each coalition $U$ consists of concave, strictly increasing, and continuous functions on $X$. Pick any $(p,w)\in P\times W.$ 
\begin{enumerate}

\item For any $x^*\in x_r(p,w)$, the point $y=x^*$ solves problem (\ref{Prb:C}) at $\bar{x}=x^*$.
\item When $r$ is skew-symmetric, the converse also holds: A point $x^*\in B(p,w)$ belongs to $x_r(p,w)$ if and only if it solves problem (\ref{Prb:C}) at $\bar{x}=x^*$. 
\item A point $x^*\in B(p,w)$ solves problem (\ref{Prb:C}) at $\bar{x}=x^*$ if and only if it maximizes $r_U(\cdot,x^*)$ on $B(p,w)$ for every $U\in \Omega$. 
\item A point $x^*\in X$ with $p x^*=w$ and $x^*\gg 0$ solves problem (\ref{Prb:C}) if and only if there exist numbers $\lambda_U >0$ for $U \in \Omega$ such that
\begin{equation}
\label{FOC}
\lambda_U p \in \partial_1 r_U(x^*,x^*)\quad \forall U \in \Omega,
\end{equation}
where $\partial_1 r_U(\cdot,x^*)$ denotes the superdifferential of the concave function $r_U(\cdot,x^*)$. If, in addition, a coalition $U$ consists of finitely-many differentiable functions, then
$\partial_1 r_U(x^*,x^*)$ equals the convex hull of the set $\{\nabla u(x^*):u\in U\}.$

\end{enumerate}
\end{lemma}

Expression (\ref{FOC}) is a Kuhn-Tucker condition for maximization of the strictly increasing, concave (coalitional) utility function $r_U(\cdot,x^*)$ at an interior point of the budget hyperplane. Since the function $r_U$ has kinks, the superdifferential $\partial_1 r_U$ takes the role of a uniquely defined gradient that appears in the more familiar, textbook version of the utility maximization problem.\footnote{For corner solutions, first-order conditions (\ref{FOC}) are extended in the usual way, with $L+1$ nonnegative multipliers and complementary slackness conditions (see Corollary 28.3.1 in \citet{Rockafellar1972}).} 

Finally, in statement (iv), the equality $\partial_1 r_U(x^*,x^*)=\operatorname{co}\{\nabla u(x^*):u\in U\}$ follows from a well-known fact on the superdifferential of the pointwise minimum of finitely-many concave functions.\footnote{Specifically, if $f_1,\ldots,f_m$ are concave and differentiable functions, each mapping $X$ to $\mathbb{R}$, then the function $f(x)\equiv \min\{f_1(x),\ldots,f_m(x)\}$ satisfies $\partial f(x)=\operatorname{co} \{\nabla f_i(x):i\in I(x)\}$, where $I(x)\equiv \{i:f_i(x)=f(x)\}$ is the set of active functions at the point $x$. In statement (iv), the functions $f_i(\cdot)$ correspond to $u(\cdot)-u(x^*)$, and each of these functions is active at $x^*$ because $u(x^*)-u(x^*)=r_U(x^*,x^*)=0$ for every $u\in U$. An analogous formula can also be established for the superdifferential of the skew-symmetric CMU function constructed in Lemma \ref{lem:CMU-skewsymmetric} provided that the component functions $u_{ij}$ are concave and differentiable on $X$ (see Section 3 in \citet{boyd2022subgradients}).}

 \subsection{Expenditure Minimization}
As in the previous subsection, let us consider an asymmetric, strictly increasing, reflexive, and continuous preference function $r$. For any $x,y \in X$, we have $r(x,y)<0$ if and only if $x\succeq _r y$ is not true. That is, the set $x^\uparrow\equiv \{y\in X:r(x,y)<0\}$ equals $\{y\in X:\neg (x\succeq_r y)\}$. Moreover, our assumptions on $r$ imply
\[
\{y\in X: y\gg x \}\subseteq \{y\in X: y\succ_r x \}\subseteq x^\uparrow.
\]
Since any neighborhood of $x$ contains a strictly greater bundle, it follows that $x$ belongs to $\operatorname{cl}x^\uparrow,$ the closure of the set $x^\uparrow$.\footnote{If we also assume that $r$ is complete, $\operatorname{cl}x^\uparrow$ becomes the same as the upper contour set $\{y\in X: y\succeq_r x\}$.}

Given a price vector $p\in P$ and a fixed bundle $x$, we denote by $h(p,x)$ the set of solutions to the following expenditure minimization problem:
\begin{equation}
\label{Prb:E}
\min py\quad\text{s.t.}\quad y\in \operatorname{cl} x^\uparrow. \tag{E}
\end{equation}
The associated expenditure
function $e(p,x)$ is defined as $e(p,x)=px^{\ast}$ for $x^{\ast}\in
h(p,x).$ Thus, we have $x\in h(p,x)$ if and only if $px=e(p,x)$.

In the classical expenditure minimization problem of consumer theory, the constraint set is defined by means of a target utility level. In problem (\ref{Prb:E}), we define the constraint set using a single alternative $x$, to capture the essence of the fixed-point problem embodied in the preference maximization problem of Section \ref{sec:Pref-Max}.

\begin{lemma}
\label{lem:expenditure}
Suppose $r$ is an asymmetric, strictly increasing, reflexive, and continuous preference function on $X$. Fix a bundle $x\in X$ and a price vector $p\in P$. Set $w_{x}=px$.  
\begin{enumerate}
\item $x\in x_r (p,w_x)$ if and only if $x\in
h(p,x)$.
\item $h(p,x)$ is nonempty.

\item If $r(x,y)$ is strictly decreasing in $y$, then $\operatorname{cl}x^\uparrow= \{y\in X: r(x,y)\leq 0\}$. Moreover, for any $y\in h(p,x)$ we have $r(x,y)= 0$.
\end{enumerate}
\end{lemma}

Statement (i) in Lemma \ref{lem:expenditure} establishes a duality between fixed points of the Walrasian solution correspondence obtained from problem (\ref{Prb:P}), and those of the Hicksian solution correspondence $h(p,\cdot)$ obtained from problem (\ref{Prb:E}). 

The additional monotonicity condition in statement (iii) holds when $r$ is skew-symmetric or if it is a CMU function. In this case, the constraint in problem (\ref{Prb:E}) becomes equivalent to the inequality $r(x,y)\leq 0$. Alternatively, one may also use the equality constraint $r(x,y) = 0$. With these functional constraints, the first-order conditions in problems (\ref{Prb:E}) and (\ref{Prb:P}) are related to each other in the same way as in the classical consumer theory.

We present the proof of Lemma \ref{lem:expenditure} in Appendix \ref{app:proofs}; it is based on fairly straightforward arguments.

Many empirical demand models are derived from functional form representations of the expenditure function. Two prominent examples are the almost ideal demand system based on the PIGLOG expenditure function (\citealt{Deaton1980}), and the NQ demand system based on the normalized quadratic expenditure function (\citealt{Diewert1988}). Given a functional form representation, empirical demand functions are obtained by applying Shepard's lemma. The relevant elasticities can then be calculated from the estimated system of demand functions. We hope that the preceding discussion may serve as a first step toward similar empirical demand models based on general preference functions.

\section{Concluding Remarks \label{sec:conclusion}}

This paper offers an Afriat-like theorem for WGARP, as well as a study of the related class of incomplete or intransitive preferences. Our main result shows that coherent and strictly monotonic CMU functions fully characterize WGARP. We build a comprehensive theory of revealed preference based on this new rationalization notion. Our findings should be helpful for practitioners of revealed preference, since, from an empirical
perspective, WGARP is significantly easier to work with than GARP. 

In the Online Appendix,  we show that the proof of our main result, Theorem \ref{thm:WGARPCharacterization}, can easily be modified to obtain a characterization of \citeauthor{samuelson_note_1938}'s
(\citeyear{samuelson_note_1938}) weak axiom of revealed preference (WARP) for single-valued demand functions. This extension is based on \citeauthor{matzkin_testing_1991-1}'s \citeyearpar{matzkin_testing_1991-1} characterization of the strong axiom of revealed preference (SARP), as opposed to Afriat's characterization of GARP. The strict concavity of the utility functions provided by \citeauthor{matzkin_testing_1991-1}'s theorem makes the corresponding CMU function strictly star-concave. This leads to a refinement of \citeauthor{kim_nontransitive-nontotal_1986-1}'s (\citeyear{kim_nontransitive-nontotal_1986-1}) characterization of WARP by strictly star-convex preference relations. In particular, our characterization of WARP guarantees continuity of the rationalizing preference relation. While our functional representations establish a surprisingly close connection between WARP and WGARP, these axioms do not appear to be as close in terms of the primitive properties of associated preference relations. Indeed, we have seen that WGARP is associated with $>$-consistency of the rationalizing preference relation, as opposed to a convexity (or star-convexity) condition.

Another related work is \citeauthor{mariotti2008kind}'s (\citeyear{mariotti2008kind}) study of the Weak Congruence axiom. This is a binary version of \citeauthor{Richter1966}'s (\citeyear{Richter1966}) Congruence axiom that characterizes rationalizable choice correspondences under the assumption that all choosable alternatives in a given menu are observed by the analyst. Since GARP does not require the latter assumption, a dataset that satisfies GARP may violate the Weak Congruence axiom. The choice environment in \citet{mariotti2008kind} lacks a dominance relation, while the Weak Congruence axiom is characterized by a generalized preference maximization condition called ``justification.'' By contrast, our rationalization notion for WGARP involves maximization --in the standard sense-- of a $>$-consistent preference relation that is possibly intransitive. To summarize, despite the similarity of questions we study, there is no direct connection between our findings and \citeauthor{mariotti2008kind}'s (\citeyear{mariotti2008kind}) work on the Weak Congruence axiom.

More recently, \citet{nishimura2017comprehensive} have shown that Afriat's theorem can be adapted to a general metric space of alternatives equipped with a dominance relation. Their modification of direct revealed preference relations also allows one to incorporate nonlinear and noncomprehensive budget sets into this theory. In the Online Appendix, we show that an analogous modification of WGARP can be characterized by coherent CMU functions that feature strictly monotonic utility functions with respect to the underlying dominance relation. Focusing on the general setup of \citet{nishimura2017comprehensive}, a further supplementary result in the Online Appendix shows that coherent and monotonic CMU functions can be used to represent any reflexive preference relation that satisfies a generalized version of the $>$-consistency property.

Finally, one important area we have not discussed is welfare. \cite{varian_nonparametric_1982} shows how to calculate measures of welfare under GARP, which is essentially based on recovering preferences between any pair of bundles from the available observations. \citet{Halevy2017} and \citet{nishimura2017comprehensive} note that Varian's method hinges on the assumption of convex preferences, and suggest alternative methods based on monotonicity assumptions, in place of convexity. As we shall see in Appendix \ref{app:Non-Convexity}, under WGARP convexity of preferences becomes a testable property. In principle, this can provide an empirical framework to compare and contrast different approaches to welfare measurement. We leave
this issue to be studied elsewhere.

% \footnote{A recent literature extends revealed preference methods to recover preferences of different types of consumer demand models. Two notable examples are \citet{Blundell2003,blundell_best_2008}, which show that it is possible to substantially enhance recovery and prediction results by combining revealed-preference theory with nonparametric estimation of Engel curves. The analyses in these papers are based on WARP, which we have shown may be problematic in a setup with more than two goods due to the issue of empty demand (\citet{Blundell2003} considers 22 goods, and \citet{blundell_best_2008} uses three goods in their empirical applications). \citet{Blundell2015} shows how the methods in \citet{Blundell2003,blundell_best_2008} can be modified to derive sharp bounds on welfare measures under SARP.}

%  \bibliographystyle{ecca}
% \bibliography{WARPpaper}

\appendix

\section*{Appendices}

\section{More on the Possibility of Empty Demand} \label{appendixA}

In this appendix, we take a closer look at the relations between WGARP, rationalizability by (quasi)concave preference functions, and the possibility of empty demand. We also discuss an alternative demand notion for incomplete preferences, and show that this alternative approach is not compatible with WGARP. 

\subsection{Convexity, Its Absence, and Implications \label{app:Non-Convexity}}

A classic theorem by \citet{shafer_nontransitive_1974} shows that a skew-symmetric, continuous, and quasiconcave\footnote{We say that a preference function $r:X\times X \mapsto \mathbb{R}$ is \emph{quasiconcave} if $r(x,y)$ is quasiconcave in $x\in X$, for each fixed $y\in X$. Concavity of a preference function is defined analogously.} preference function induces a nonempty-valued demand correspondence. \citet{john_concave_2001} describes a more refined class of preference functions that satisfy a stronger concavity property, as well as strict monotonicity. Next, we present a streamlined version of this particular class of functions. 

\begin{example}\label{exp:concave} (\citet{john_concave_2001}) Fix a natural number $n$, and let $N=\{1,\ldots,n\}$. Pick a matrix $A=[a_{ij}]\in \mathbb{R}^{n\times n}$ and a set of $(L+1)$ dimensional vectors $\{(z^i,d^i):i\in N\}$ such that for every $i,j\in N$, $a_{ij}=-a_{ji}$, $ z^i\in \mathbb{R}^L_{++}$, and $d^i\in \mathbb{R}$. Consider a preference function $r:X\times X \mapsto \mathbb{R}$ defined by the following rule:
\[
r(x,y)=\min_{\lambda \in \Delta}\max_{\mu \in \Delta} \sum_{i \in N}\sum_{j \in N}\lambda_i\mu_j \left( a_{ij}+(z^ix-d^i)-(z^jy-d^j)\right),
\]
where $\Delta$ denotes the $(n-1)$ dimensional simplex. It is easily checked that, for any $x,y\in X,$
\begin{align}
r(x,y)  &=\min_{\lambda \in \Delta} \sum_{i \in N}\lambda_i (z^ix-d^i)+c_y(\lambda),\label{eq:variational}\\
 \text{where} \quad c_y(\lambda) & \equiv\max_{\mu \in \Delta} \sum_{i \in N}\sum_{j \in N}\lambda_i\mu_j (a_{ij}-(z^jy-d^j))\quad \forall\lambda \in\Delta.  \nonumber
\end{align}

Since each $z^i$ is a strictly positive vector, from equation (\ref{eq:variational}) it follows that $r(x,y)$ equals the minimum of a collection of strictly increasing and affine functions of $x$. As such, $r(x,y)$ is concave and strictly increasing in $x,$ for each fixed $y\in X$. Moreover, \citet{john_concave_2001} shows that $r$ is skew-symmetric, as a result of the minimax theorem and the negative symmetry property $a_{ij}=-a_{ji}.$

\end{example}

Since they are skew-symmetric, strictly increasing and continuous, preference functions described in Example \ref{exp:concave} satisfy all conditions in statement (i) of Theorem \ref{thm:WGARPCharacterization}, while their concavity guarantees a nonempty-valued demand correspondence.\footnote{In the terminology of \citet{MMR2006}, expression (\ref{eq:variational}) defines a \emph{variational} function in $x$, for each fixed $y$.} \citet{john_concave_2001} also shows that a dataset $O^{T}=\{(p^{t},x^{t})\}_{t\in\mathbb{T}}$ can be rationalized by such a preference function if and only if there exist a set of numbers $\{\pi^t:t\in\mathbb{T}\} \subset \mathbb{R}_{++}$ such that 
\begin{equation}
\label{eq:John}
(\pi^sp^s-\pi^tp^t)(x^s-x^t)\leq 0,
\end{equation}
for every $s,t\in \mathbb{T}$. If WGARP fails, we have $p^s(x^s-x^t)>0$ and $-p^t(x^s-x^t)\geq0$ for some $s,t\in \mathbb{T}$, which makes the inequality (\ref{eq:John}) impossible for any choice of strictly positive numbers, $\pi^s,\pi^t$. Thus, \citet{john_concave_2001}'s axiom implies WGARP. However, the existential clause in \citet{john_concave_2001}'s axiom makes it less intuitive than WGARP, and may create difficulties in implementations. It is also not known whether \citet{john_concave_2001}'s axiom can be weakened toward a characterization of rationalizability by quasiconcave preference functions. This rationalizability problem and its relation to CMU representations are nontrivial questions, which we leave for future research.

Let us show here that, unlike Afriat's theorem, rationalizability by a quasiconcave preference function is not implied by WGARP. That is, it is not always possible to select a quasiconcave preference function in statement (i) of Theorem \ref{thm:WGARPCharacterization}.

\begin{example}\label{ExampleC}
Following Example \ref{ex:DemandCounterfactuals}, consider the dataset $O^{3}$ with prices $p^{1}=(4,1,5)$, $p^{2}=(5,4,1)$,
$p^{3}=(1,5,4)$, and bundles $x^{1}=(4,1,1)$, $x^{2}=(1,4,1)$,
$x^{3}=(1,1,4)$. As noted earlier, this dataset satisfies
WGARP.

We claim that $O^3$ cannot be rationalized by an asymmetric, strictly increasing, and quasiconcave preference function $r:X\times X \mapsto \mathbb{R}$. Suppose by contradiction that a rationalizing function $r$ satisfies these properties. Define a new bundle
\[
\hat{x}=\frac{1}{3}(x^{1}+x^{2}+x^{3})=(2,2,2).
\]
Note that $p^{t}(x^{t}-\hat{x})=2>0$ for all $t=1,2,3$. Hence, there exists a bundle $\hat{z}$ such that $\hat{z}\gg \hat{x}$, and $p^{t}(x^{t}-\hat{z})>0$ for $t=1,2,3$. So, rationalization requires $r(x^{t},\hat{z})\geq0$ for $t=1,2,3$. Since $\hat{x}$ is a convex combination of the observed bundles, from quasiconcavity of $r$ it follows that $r(\hat{x},\hat{z})\geq \min_{t=1,2,3}\{r(x^{t},\hat{z})\}\geq0$. Then by strict monotonicity of $r$, we must also have $r(\hat{z},\hat{z})>0$, in contradiction to the asymmetry requirement. 

%This implies that for at least one of the three observed %bundles $x^{t}$ for $t=1,2,3$, it is not the case that %$x^{t}$ is preferred to $x^{T+1}$, 
 %i.e., $r(x^t, x^{T+1})<0$ for at least one $t=1,2,3$. %However, note that, for all $t=1,2,3$, we have $p^{t}(x^{t}-x^{T+1})=2>0$.
%Thus, by the rationalization of Theorem 1,
 %all observed commodity bundles must be weakly  %preferred to
%the new bundle, i.e., $r(x^{t},x^{T+1})\geq 0$ for all %$t=1,2,3$, a contradiction. Hence,
%the dataset $O^{3}$ cannot be rationalized
%by a strictly quasiconcave and asymmetric (or of course, skew-symmetric) preference function. Also, it follows that it cannot be rationalized by a strictly concave and skew-symmetric preference function either, akin to the Shafer model of a concave nontransitive consumer (\cite{john_concave_2001}).\footnote{One way to shed light on this failure is that the extended dataset, adding the fourth bundle, violates WGARP. 
%The question is whether there exists a price-vector $p^{T+1}$ such
%that the extended dataset $O^{3}\cup(p^{T+1},x^{T+1})$ satisfies
%WGARP. 
%We know $p^{t}(x^{t}-x^{T+1})=2>0$ for all $t=1,2,3$,
%and we also know that there is no $p\in P$ such that $p(x^{T+1}-x^{t})<0$
%simultaneously for all $t=1,2,3$, since adding up these inequalities leads to a contradiction.}

\end{example}

Paul Samuelson's ``eternal darkness'' conjecture asserts that nonconvexity of preferences cannot be detected in a finite dataset (see \citet{Pol-Ren2016} and references therein). While this is certainly true under GARP, Example \ref{ExampleC} gives the opposite message in terms of the properties of preference functions that are compatible with WGARP.\footnote{Specifically, in Example \ref{ExampleC}, if a rationalizing preference function $r$ is strictly increasing and asymmetric, then $\succeq_r$ is necessarily a non-convex preference relation because $\{y\in X: y\succeq_r \hat{z}\}$ is not a convex set.}  

A related conjecture by \citet{kihlstrom_demand_1976-1} asserts that a continuously differentiable (and single-valued) demand function $x(p,w)$ can be rationalized by a strictly convex preference relation whenever WARP holds on the grand set $\{x(p,w):(p,w) \in P\times W\}$. Example \ref{ExampleC} also negates a finite version of this conjecture. Indeed, the rationalization condition implies $x^t\succsim_r \hat{x}$ for $t=1,2,3$. Since $\hat{x}$ sits in the middle of the pairwise distinct points $x^1,x^2,x^3$, strict convexity of $\succsim_r$ would imply $ \hat{x}\succ_r \hat{x}$, an absurdity.  

In Example \ref{ExampleC}, the absence of a rationalizing quasiconcave function is accompanied by a demand-existence problem. All asymmetric and strictly increasing preference functions that rationalize the dataset $O^3$ yield an empty demand correspondence at the price-wealth pair $(\hat{p},\hat{w})=\frac{1}{3}\sum_{i=1}^{3}(p^{i},w^{i})$. In fact, as we have seen in Example \ref{ex:DemandCounterfactuals}, it becomes impossible to make a demand prediction for the out-of-sample price-wealth pair $(\hat{p},\hat{w})$, without violating WGARP.

The absence of a preference maximizing bundle may reflect the agent's indecision. Next, we will see that dropping the preference maximization condition leads to alternative difficulties for the purposes of demand analysis.

\subsection{Incompleteness and WGARP}
\label{app:Incomplete}

%Indecision may be viewed as the root cause of the non-existence of a preference maximizer on a budget set. 

In the context of boundedly rational behavior, it sometimes becomes fruitful to distinguish between agents' observable choices and their psychological preferences based on welfare judgments. In particular, it has been noted that a consumer with a transitive but incomplete psychological preference relation may exhibit a cyclic behavior when she is forced to make choices (\citet{Man2006}, \citet{Costa-Gomes2022}). While a complete and intransitive preference function may be used to model the observed behavior, it is also possible to use an incomplete preference relation and directly model the underlying psychological preferences.\footnote{In this paper, we follow the former revealed-preference approach. Indeed, in Theorem \ref{thm:WGARPCharacterization}, we have seen that the completeness assumption causes no loss of generality for our purposes. But our approach is also compatible with intermediate cases such as a supermajority voting rule, which is a behavioral rule that is more decisive than the underlying Pareto dominance relation, yet not fully complete.} In the latter approach, one may also wish to modify the definition of demand, and recognize choice possibility for any undominated alternative. Specifically, given an incomplete preference function $r$, one may define an alternative demand correspondence by the rule 
\[
x^{+}_r(p,w)\equiv \{x\in B(p,w):\neg (y\succ_r x), \; \forall y \in B(p,w)\}.
\]

Suppose $r$ is a continuous, transitive, but possibly incomplete preference function. Then, by Theorem 1 in \citet{Evr-Ok2011}, there exists a nonempty set $U\subseteq C(X)$ such that
\begin{equation}
\label{incomplete MU}
x \succeq_r y\quad \text{if and only if}\quad \inf_{u\in U} (u(x)-u(y))\geq 0,
\end{equation}
for every $x,y \in X$. 

On the right-hand side of (\ref{incomplete MU}), we can identify the set $U$ with a degenerate coalition structure $\Omega''\equiv\{U\}$. Although this is a coherent coalition structure, the more inclusive demand notion $x^{+}_r(p,w)$ will typically generate violations of WGARP. This is the content of the next observation.  

\begin{lemma}
\label{lem:No-WGARP2}
In representation (\ref{incomplete MU}), suppose every function $u$ in the set $U$ is strictly quasiconcave, continuous, and strictly increasing on $X$. Then one of the following statements holds true: 
\begin{enumerate}
\item \label{lem:No-WGARP(i)} Any dataset $O^T=\{(p^t,x^t)\}_{t\in\mathbb{T}}$ satisfies GARP provided that $x^t\in x^{+}_r(p^t,w^t)$ for every $t\in\mathbb{T}$.
\item \label{lem:No-WGARP(ii)} There exist price-wealth pairs $(p^1,w^1),\;(p^2,w^2)$ in $P\times W$ and bundles $x^1,\; x^2$ such that $x^1\in x^{+}_r(p^1,w^1), \; x^2\in x^{+}_r(p^2,w^2),\; p^1x^2<w^1,$ and $p^2x^1<w^2$.
\end{enumerate}
\end{lemma}

From a technical standpoint Lemmas \ref{lem:No-WGARP} and \ref{lem:No-WGARP2} are very similar.\footnote{Technically, the only difference between Lemmas \ref{lem:No-WGARP} and \ref{lem:No-WGARP2} is the strict quasiconcavity condition in the latter, which guarantees that any maximizer of a function $u\in U$ on a budget set $B(p,w)$ belongs to $x^{+}_r(p,w).$} The focus of Lemma \ref{lem:No-WGARP2} is on the alternative demand correspondence $x^{+}_r(\cdot,\cdot)$, which also proves incompatible with WGARP. This rules out another reasonable approach that could deliver a nonempty-valued demand correspondence.

It should also be noted that in the present framework, the justifiable CMU model in Section \ref{Sec:Quasitransitive} can rationalize any finite dataset, and the same holds for representation (\ref{incomplete MU}) coupled with the demand correspondence $x^{+}_r(\cdot,\cdot)$. More specifically, these representations gain empirical content only if we observe \emph{all} choosable alternatives from a given budget set, as opposed to the ``weak'' rationalization notion $x^t \in x_r(p^t,w^t)$ that we adopt in this paper. Hence, shortly put, relying on these preference classes in order to solve the demand-existence problem poses alternative testability and prediction problems, in addition to systematic violations of WGARP.\footnote{In most real-life situations, we cannot aspire to observe but a subset of a consumer's \emph{true} demand correspondence. Accordingly, the weak rationalization approach of \citet{afriat_construction_1967} and \citet{varian_nonparametric_1982}, which we follow in this paper, does not attempt to uniquely identify the consumer's demand correspondence, although it allows for the existence of other bundles that are indifferent to those observed in the dataset. Following a different path, \citet{balakrishnan2022inference} study how the choice correspondence of an agent can be inferred from a history of repeated choices. Statistical techniques concerned with multiple-hypotheses testing problems play a central role in their analysis. Alternatively, one can think of special experimental designs that allow subjects to select multiple alternatives at a time, so that the analyst can estimate their choice correspondence (e.g. \citet{bouacida2021identifying}).}

\section{Omitted Proofs}\label{app:proofs}
%NEW PROOFS:

For the proof of Lemma \ref{lem:pref_max} the reader may refer to the related arguments in Section \ref{sec:Pref-Max}. In this appendix, we prove the remaining results. 

The following claim will allow us to appeal to Berge's maximum theorem in relation to CMU functions. In what follows, $\mathcal{K}$ denotes the collection of nonempty compact subsets of $C(X)$ equipped with the Hausdorff metric $d_H$ induced by the compact convergence metric $d$ on $C(X)$.

\begin{claim}
\label{Claim 1}
Define a correspondence $\psi:\mathcal{K}\twoheadrightarrow C(X)$ by the rule $\psi(U)=U$ for $U\in \mathcal{K}$, which maps elements of the metric space $(\mathcal{K}, d_H)$ into subsets of $(C(X),d)$. The correspondence $\psi$ is both upper and lower hemicontinuous.

\begin{proof} The distance between a point $u\in C(X)$ and a set $V\subseteq C(X)$ is given by 
$d(u,V)\equiv\inf_{v\in V}d(u,v)$. The Hausdorff distance between two sets $U,V\in \mathcal{K}$ satisfies 
\[d_H(U,V)=\max\{\max_{u\in  U}d(u,V),\max_{v\in V}d(v,U)\}.\] 

Fix any $U \in \mathcal{K}$. Consider an open set $O\subseteq C(X)$ such that $\psi(U)\subseteq O$. For upper hemicontinuity, we need to show that there exists an $\varepsilon>0$ such that $d_H(U,V)<\varepsilon$ implies $\psi(V)\subseteq O$, for every $V\in \mathcal{K}$. 

Without loss of generality, suppose $O^c\equiv C(X)\setminus O$ is nonempty. Since $U=\psi(U)$ does not intersect the closed set $O^c$, we have $d(u,O^c)>0$ for every $u\in U$. As $U$ is compact and $d(\cdot,O^c)$ is a continuous function, there exists an $\varepsilon >0$ such that $d(u,O^c)>\varepsilon$ for every $u\in U$. This implies that the set $B_\varepsilon(U)\equiv \bigcup_{u\in U} \{v\in C(X): d(u,v)<\varepsilon\}$ is contained in $O$. Moreover, for any $V\in \mathcal{K}$ with $d_H(U,V)<\varepsilon$, we have $d(v,U)<\varepsilon$ for every $v\in V$, which implies $V\subseteq B_\varepsilon(U)$. Then $\psi(V)=V$ is also contained in $O$, as we seek.

To prove lower hemicontinuity, fix an open set $O \subseteq C(X)$ such that $\psi(U)\cap O \neq \emptyset.$ We need to show that there exists an $\varepsilon>0$ such that $d_H(U,V)<\varepsilon$ implies $\psi(V)\cap O\neq \emptyset$, for every $V\in \mathcal{K}$.

Pick any point $u$ from the nonempty set $\psi(U)\cap O$. Since $O$ is an open set, there exists an $\varepsilon>0$ such that $B_\varepsilon(u)$ is contained in $O$. Moreover, since $u$ belongs to $U=\psi(U)$, we have $d(u,V)\leq d_H(U,V)$ for any $V\in \mathcal{K}$. Thus, $d_H(U,V)<\varepsilon$ implies $d(u,V)<\varepsilon$, which means $d(u,v)<\varepsilon$ for some $v\in V$. Then $v$ belongs to $V\cap B_{\varepsilon}(u)$, which is a subset of $V\cap O$. This shows that the sets $\psi(V)=V$ and $O$ have nonempty intersection for any $V\in \mathcal{K}$ with $d_H(U,V)<\varepsilon$. 

\end{proof}
\end{claim}

\subsection*{Proof of Lemma \ref{Lem:Basic CMU}}

\subsubsection*{(i)} Let $\mathcal{K}'\equiv \mathcal{K}\times X\times X.$ Define a function $f: \mathcal{K}'\times C(X)\mapsto \mathbb{R}$ and a correspondence $\psi:\mathcal{K}'\twoheadrightarrow C(X)$ as 
\[
f(U,x,y,u)=u(x)-u(y) \quad \text{and}\quad  \psi(U,x,y)=U,
\]
for every $(U,x,y)\in \mathcal{K}'$ and $u\in C(X)$. 

Since $\psi(U,x,y)$ does not depend $x$ or $y$, Claim \ref{Claim 1}
implies that the correspondence $\psi$ is upper and lower hemicontinuous.

Next, we show that the function $f$ is continuous. Consider a sequence $\{(U_n,x_n,y_n,u_n)\} $ that converges to $(U,x,y,u)$ in $\mathcal{K}'\times C(X)$. There exists a compact set $B\subset X$ that contains the sequences $\{x_n\},\{y_n\}$ and their limits $x,y$. Moreover, by compact convergence criterion, $u_n(z)$ converges to $u(z)$ uniformly for $z\in B$, as $n\rightarrow \infty $. Thus, for any $\varepsilon>0$, there exists an $\bar{n}\in \mathbb{N}$ such that $|u_n(x_n)-u(x_n)|<\varepsilon$ and $|u_n(y_n)-u(y_n)|<\varepsilon$ for every $n\geq \bar{n}$. By continuity of $u$, we also have $|u(x_n)-u(x)|<\varepsilon$ and $|u(y_n)-u(y)|<\varepsilon$ for large indices $n$. From the triangle inequality, it follows that $|u_n(x_n)-u(x)|$ and $|u_n(y_n)-u(y)|$ are both less than $2\varepsilon$, for all sufficiently large $n$. This yields $|f(U_n,x_n,y_n,u_n)-f(U,x,y,u)|=|u_n(x_n)-u(x)+u(y)-u_n(y_n)|<4\varepsilon$ for large $n$. Hence, $f$ is a continuous function on $\mathcal{K}'\times C(X)$, as we claimed.   

From Berge's maximum theorem \cite[Theorem 17.31]{Alip-Border2006}, it follows that $f(U,x,y,u)$
attains its minimum over $u\in \psi(U,x,y)$, and $r(U,x,y)\equiv \min_{u\in \psi(U,x,y)}f(U,x,y,u)$ is a continuous function\footnote{Lemma 5 in \citet{nishimura2016utility} establishes continuity of the function $f$ with slightly different arguments, assuming a compact metric space of alternatives $X$.} of  the parameters $(U,x,y)\in \mathcal{K}'.$

Now, define a further correspondence $\psi_1:X \times X \twoheadrightarrow \mathcal{K}$ by the rule $\psi_1(x,y)=\Omega$ for every $(x,y) \in X \times X$. Being a constant correspondence with a nonempty and compact image set, $\Omega$, the correspondence $\psi_1$ also satisfies the conditions of Berge's theorem. Hence, the value function $r^\Omega(x,y) =\max_{U\in \psi_1(x,y)} r(U,x,y)$ is continuous in $(x,y)\in X\times X$.

\subsubsection*{(ii)} Given a collection $\mathcal{G}$ of strictly
increasing (respectively, decreasing), real valued functions on $X$, the functions $\max_{g\in
\mathcal{G}}g(x)$ and $\min_{g\in\mathcal{G}}g(x)$, if well-defined, are also
strictly increasing (respectively, decreasing) on $X$.

In the context of a CMU function $r^\Omega(x,y)$, Weierstrass theorem ensures that the functions $r(U,x,y)\equiv \min_{u\in U}u(x)-u(y)$ for $U\in \Omega$, and $r^{\Omega}(x,y)=\max_{U\in\Omega}r(U,x,y)$ are well-defined. 

If a function $u\in C(X)$ is strictly increasing, then $x\mapsto u(x)-u(y)$ strictly increases in $x$, while $y\mapsto u(x)-u(y)$ strictly decreases in $y$. These properties are inherited by $r(U,x,y)$ for every $U\in \Omega$, and consequently, also by $r^{\Omega}(x,y)$ provided that every coalition $U$ consists of strictly increasing functions.    

\subsection*{Proof of Lemma \ref{Lem:Asymmetry}}

Consider a strictly increasing and asymmetric preference function $r$. Let us first show that $\succeq_r$ is strictly $>$-consistent in the following sense: For any $x,y,z\in X$, we have $x\succ_r y$ whenever $x> z$ and $z\succeq_r y$. Indeed, $x> z$ implies $r(x,y)>r(z,y)$ by strict monotonicity of $r$, while $z\succeq_r y$ means $r(z,y)\geq 0$. Then we get $r(x,y)>0$, and also $r(y,x)<0$ by asymmetry of $r$. These inequalities jointly imply $x\succ_r y$.

Now, suppose that the preference function $r$ rationalizes a dataset $O^T$. Pick any pair of observations $s,t\in\mathbb{T}$. By rationalization, we have $x^{s}\in x_r (p^s,w^s)$. Hence, if $p^{s}x^{s}>p^{s}x^{t}$, then Claim \ref{budget interior} below implies $x^{s}\succ_r x^t$. By rationalization, we also know that $x^{t}\succeq_r y$ for every $y \in B(p^t,w^t)$. It follows that $x^{s}$ does not belong to $B(p^t,w^t)$, that is, $p^{t}x^{s}>p^{t}x^{t}$. This verifies WGARP.   

\begin{claim}\label{budget interior} Consider a strictly increasing and asymmetric preference function $r$. For any $(p,w)\in P\times W$, $x\in x_r(p,w)$ and $z\in X$ with $pz<w$, we have $x\succ_r z$.
\end{claim}

\begin{proof} By definitions, $x\in x_r(p,w)$ and $pz<w$ imply $x\succeq_r z$. Thus, we have $x\succ_r z$ if and only if $z\succeq_r x$ is not true. Suppose by contradiction that $z\succeq_r x$. Since $pz<w$, there exists an $x'\in X$ such that $x'\gg z$ and $px'<w$. But $x'\gg z$ and $z\succeq_r x$ jointly imply $x'\succ_r x$ because $\succeq_r$ is strictly $>$-consistent. This contradicts the hypothesis $x\in x_r(p,w)$. 
\end{proof}

\subsection*{Proof of Lemma \ref{lem:coherent}}

Let $r^\Omega$ be a coherent CMU function. Pick any $x,y \in X$ such that $r^\Omega(x,y)>0$. This means that there exists a $\hat{U}\in \Omega$ such that 
$u(x)-u(y)>0$ for every for every $u\in\hat{U}$. Then, by coherence, for any $U\in\Omega$,
there is a $\hat{u}\in U\cap\hat{U}$ such that $\hat{u}(y)-\hat{u}(x)<0$.
This implies $r^\Omega (y,x)=\max_{U\in\Omega}\min_{u\in U}(u(y)-u(x))<0$, as we wanted to show.

From similar arguments it also follows that $r^\Omega (x,y)\geq0$ implies $r^\Omega(y,x)\leq0$.

\subsection*{Proof of Lemma \ref{lem:CMU-skewsymmetric}}
Define a function $f:\Delta\times\Delta\mapsto C(X)$ as $f(\lambda,\mu)=u_{\lambda,\mu}= \sum_{i\in N}\sum_{j\in N}\lambda_{i}\mu_{j}u_{ij}$ for $(\lambda,\mu) \in \Delta\times\Delta$. Since $C(X)$ equipped with the compact convergence metric $d$ is a topological vector space, for each $i,j\in\{1,\dots,n\}$ the function $(\lambda,\mu)\mapsto \lambda_{i}\mu_{j}u_{ij}$ is continuous on $\Delta\times\Delta$. The sum of these functions, $f$, is also continuous.

Recall that the image of a compact set under a continuous function is also compact. Thus, $U_\lambda= \{u_{\lambda,\mu}:\mu \in \Delta\}=\{f(\lambda,\mu):\mu \in \Delta\}$ is a compact subset of $C(X)$ for every $\lambda\in \Delta$.

\begin{claim}\label{cl:Con_F}
$F:\lambda\mapsto U_\lambda$ is continuous as a function that maps $\Delta$ into $(\mathcal{K},d_H)$.
\end{claim}

\begin{proof} Pick a convergent sequence $\{\lambda^m\}$ in $\Delta$, and set $\lambda\equiv \lim_m\lambda^m$. For any $u\in C(X)$, let $\mathcal{O}_u$ denote the collection of all open subsets of $C(X)$ that contain $u$. We denote by $\operatorname{Ls}U_{\lambda^{m}}$ the set of all points $u\in C(X)$ such that for every $O\in \mathcal{O}_u$, we have $O\cap U_{\lambda^{m}}\neq \emptyset$ for infinitely many $m\in \mathbb{N}$. Also, $\operatorname{Li}U_{\lambda^{m}}$ denotes the set of all points $u\in C(X)$ such that for every $O\in \mathcal{O}_u$, we have $O\cap U_{\lambda^{m}}\neq \emptyset$ for all but finitely many indices $m$. 

We need to show that the sequence $\{U_{\lambda^{m}}\}$ converges to $U_\lambda$ in Hausdorff distance, $d_H$. For every $m\in \mathbb{N}$, the set $U_{\lambda^{m}}$ is a nonempty, compact subset of the compact set $\{f(\lambda,\mu):(\lambda,\mu)\in \Delta \times \Delta\}$. Thus, according to a well-known characterization, $\{U_{\lambda^{m}}\}$ converges to $U_{\lambda}$ with respect to $d_H$ if and only if $U_\lambda=\operatorname{Li}U_{\lambda^{m}}=\operatorname{Ls}U_{\lambda^{m}}$; see Theorem 3.93 in \cite{Alip-Border2006}. Note that $\operatorname{Li}U_{\lambda^{m}}\subseteq\operatorname{Ls}U_{\lambda^{m}}$. Thus, it suffices to show that  
\[
U_\lambda\subseteq\operatorname{Li}U_{\lambda^{m}} \quad \text{and} \quad \operatorname{Ls}U_{\lambda^{m}}\subseteq U_\lambda.
\]

To prove the former inclusion, pick any $u\in U_{\lambda}$. Then $u=u_{\lambda,\mu}$ for some $\mu\in \Delta$. By continuity of $f$, we have $u_{\lambda,\mu}=f(\lambda,\mu)=\lim_m f(\lambda^m,\mu)=\lim_m u_{\lambda^m,\mu}$. Thus, for any open set $O\in\mathcal{O}_u$, there exists an $\bar{m}\in \mathbb{N}$ such that $m\geq \bar{m}$ implies $u_{\lambda^m,\mu}\in O$. Since $u_{\lambda^m,\mu}$ belongs to $U_{\lambda^m}$, it follows that $O\cap U_{\lambda^{m}}\neq \emptyset$ for every $m\geq \bar{m}$. Hence, $u_{\lambda,\mu}$ belongs to $\operatorname{Li}U_{\lambda^{m}}$, as we seek.

Finally, pick any $u\in \operatorname{Ls}U_{\lambda^{m}}$. Then, by definition of $\operatorname{Ls}U_{\lambda^{m}}$, we can extract a subsequence of indices $\{\lambda^{m_l}\}$ and a sequence $\{u^l\}$ such that $\lim_l u^l=u$, and $u^l\in U_{\lambda^{m_l}}$ for every $l\in \mathbb{N}$. Since $u^l$ belongs to $U_{\lambda^{m_l}}$, for every $l\in \mathbb{N}$ there exists a $\mu^{m_l}\in \Delta$ such that $u^l=u_{\lambda^{m_l},\mu^{m_l}}$. As the set $\Delta\times\Delta$ is compact, $\{(\lambda^{m_l},\mu^{m_l})\}$ has a convergent subsequence. To simplify the notation, suppose the sequence $\{(\lambda^{m_l},\mu^{m_l})\}$ itself converges as $l\rightarrow\infty$, and let $(\lambda,\mu)$ be its limit. Continuity of the function $f$ implies   $u_{\lambda,\mu}=\lim_lu^l$. Since $C(X)$ is a metric space, the convergent sequence $\{u^l\}$ has a unique limit point. It follows that $u=u_{\lambda,\mu}$, and hence, $u$ is an element of $U_{\lambda}$. This proves the inclusion $\operatorname{Ls}U_{\lambda^{m}}\subseteq U_\lambda$.  
\end{proof}

Since the function $F:\lambda\mapsto U_\lambda$ is continuous and $\Delta$ is a compact set, the image $\{F(\lambda):\lambda\in \Delta \}=\Omega(\mathcal{M})$ is a compact subset of $\mathcal{K}$.

As noted in Section \ref{Sec: WGARP-Compatible CMU}, the coalition structure $\Omega(\mathcal{M})$ is coherent because $u_{\lambda,\hat{\lambda}}=u_{\hat{\lambda},\lambda}$ for every $\lambda,\hat{\lambda}\in \Delta$.

It remains to show that $r^\Omega(\mathcal{M})$ is skew-symmetric. Let us write $r$ in place of $r^\Omega(\mathcal{M})$. Fix any $x,y \in X$, and set $r_{ij}(x,y)=u_{ij}(x)-u_{ij}(y)$ for every $i,j\in N$. Note that
\[
r(x,y)  =\max_{\lambda\in\Delta}\min_{\mu\in\Delta}\sum_{i\in N}\sum_{j\in N}\lambda_{i}\mu_{j}r_{ij}(x,y)=\min_{\mu\in\Delta}\max_{\lambda\in\Delta}\sum_{i\in N}\sum_{j\in N}\lambda_{i}\mu_{j}r_{ij}(x,y).
\] 
Here, the first equality follows from the definitions, and the second one from von Neumann's classical minimax
theorem. 

Since $-r_{ij}(x,y)=u_{ij}(y)-u_{ij}(x)=r_{ij}(y,x)$, we also obtain
\begin{align*}
-r(x,y) & =-\min_{\mu\in\Delta}\max_{\lambda\in\Delta}\sum_{i\in N}\sum_{j\in N}\lambda_{i}\mu_{j}r_{ij}(x,y)\\
&=\max_{\mu\in\Delta}\min_{\lambda\in\Delta}\sum_{i\in N}\sum_{j\in N}\lambda_{i}\mu_{j}(-r_{ij}(x,y))
=\max_{\mu\in\Delta}\min_{\lambda\in\Delta}\sum_{i\in N}\sum_{j\in N}\lambda_{i}\mu_{j}r_{ij}(y,x).
\end{align*}

Moreover, the symmetry assumption $u_{ij}=u_{ji}$ implies $r_{ij}(y,x)=r_{ji}(y,x)$ for every $i,j\in N$. It follows that 
\[
-r(x,y)  =\max_{\mu\in\Delta}\min_{\lambda\in\Delta}\sum_{i\in N}\sum_{j\in N}\lambda_{i}\mu_{j}r_{ij}(y,x)=\max_{\mu\in\Delta}\min_{\lambda\in\Delta}\sum_{i\in N}\sum_{j\in N}\lambda_{i}\mu_{j}r_{ji}(y,x).
\]
The last step is to swap the variables $\lambda$ and $\mu$, and also the indices $i$ and $j$:
\begin{align*}
\max_{\mu\in\Delta}\min_{\lambda\in\Delta}\sum_{i\in N}\sum_{j\in N}\lambda_{i}\mu_{j}r_{ji}(y,x)&=\max_{\lambda\in\Delta}\min_{\mu\in\Delta}\sum_{i\in N}\sum_{j\in N}\mu_{i}\lambda_{j}r_{ji}(y,x)\\
&=\max_{\lambda\in\Delta}\min_{\mu\in\Delta}\sum_{i\in N}\sum_{j\in N}\lambda_{i}\mu_{j}r_{ij}(y,x)=r(y,x).
\end{align*}
We conclude that $-r(x,y)=r(y,x)$.

\subsection*{Proof of Theorem \ref{thm:WGARPCharacterization}}

The implications (iii) $\Longrightarrow$ (ii), and (ii) $\Longrightarrow$ (i) are obvious, while Lemma \ref{Lem:Asymmetry} shows that (i) implies (iv). Here, we focus on the remaining implications. We also provide a separate proof for the implication (iv) $\Longrightarrow$ (ii) that features a finite coalition structure.

\subsubsection*{(iv) $\protect\Longrightarrow$ (iii)}

Suppose that the dataset $O^{T}=\{(p^t,x^t)\}_{t\in \mathbb{T}}$ satisfies WGARP. Without loss of generality, assume that $T\geq 3$. Let $\mathcal {T}_2$ denote the collection of nonempty subsets of $\mathbb{T}$ that have $2$ elements. For every $\mathbb{S}\in \mathcal{T}_2$, we denote by $O^2_\mathbb{S}$ the sample consisting
of the two observations $\{(p^s,x^s)\}_{s\in\mathbb{S}}$. Being a subset of $O^{T}$, the dataset $O^2_\mathbb{S}$ satisfies WGARP. Since $O^2_\mathbb{S}$ has only two observations, it also satisfies GARP. Thus, by Afriat's theorem, $O^2_\mathbb{S}$ is rationalized by a continuous, concave, and strictly increasing function $u_\mathbb{S}:X\mapsto \mathbb{R}$, so that $u(x^s)\geq u(y)$ for every $y \in B(p^s,w^s)$ and $s \in \mathbb{S}$. 

For every ordered pair $(s,t)\in\mathbb{T}\times\mathbb{T}$ with $s\neq t$, define a function $u_{st}$ as $u_{st}=u_{\{s,t\}}$. Also, for every $t\in \mathbb{T}$, we select an $s\in \mathbb{T}$ that is distinct from $t$, and let $u_{tt}=u_{\{s,t\}}$. Then the $T\times T$ matrix $\mathcal{M}\equiv[u_{st}]$ is symmetric, that is, $u_{st}=u_{ts}$ for every $s,t\in\mathbb{T}$. We define the associated coalition structure $\Omega(\mathcal{M})$ as in equation (\ref{def:Omega(M)}), with $n\equiv T$. By Lemmas \ref{Lem:Basic CMU} and \ref{lem:CMU-skewsymmetric}, this coalition structure induces a continuous, skew-symmetric, and coherent CMU function $r^{\Omega(\mathcal{M})}$. 

It remains to show that $r^{\Omega(\mathcal{M})}$ rationalizes the dataset $O^{T}$ in the sense of Definition \ref{PrefRat}. Fix an index $\bar{s}\in \mathbb{T}$. A generic coalition $U_\lambda$ in $\Omega(\mathcal{M})$ is defined as $U_\lambda =\{\sum_{s\in \mathbb{T}}\sum_{t\in \mathbb{T}}\lambda_{s}\mu_{t}u_{st}:\mu \in \Delta\}$, for $\lambda \in \Delta$. With $\lambda=\bold{1}_{\bar{s}}$, we have $\lambda_{s}=0$
if $s\neq \bar{s}$ and $\lambda_{s}=1$ if $s=\bar{s}$, which gives a coalition $U_{\bar{s}} =\{\sum_{t\in \mathbb{T}}\mu_{t}u_{\bar{s}t}:\mu \in \Delta\}$. Pick any $y\in B(p^{\bar{s}},w^{\bar{s}})$. By construction, $x^{\bar{s}}$ maximizes $u_{\bar{s}t}$ on $B(p^{\bar{s}},w^{\bar{s}})$ for every $t\in \mathbb{T}$. Hence, we have $u_{\bar{s}t}(x^{\bar{s}})\geq u_{\bar{s}t}(y)$ for every $t\in \mathbb{T}$. This implies  $\sum_{t\in \mathbb{T}}\mu_{t}u_{\bar{s}t}(x^{\bar{s}})\geq \sum_{t\in \mathbb{T}}\mu_{t}u_{\bar{s}t}(y)$ for every $\mu \in \Delta$, and hence, $\min_{u\in U_{\bar{s}}}\left((u(x^{\bar{s}})-u(y)\right)\geq 0$. Since $U_{\bar{s}}$ belongs to $\Omega(\mathcal{M})$, we conclude that 
\[
r^{\Omega(\mathcal{M})}(x^{\bar{s}},y)=\max_{U\in\Omega(\mathcal{M})} \min_{u\in U} (u(x^{\bar{s}})-u(y))\geq 0. 
\]
This completes the proof of rationalization because $\bar{s}\in \mathbb{T}$ and $y\in B(p^{\bar{s}},w^{\bar{s}})$ are arbitrarily selected.

\subsubsection*{(iv) $\protect\Longrightarrow$ (ii)}
Suppose that the dataset $O^{T}=\{(p^t,x^t)\}_{t\in \mathbb{T}}$ satisfies WGARP. As in the proof of statement (iii) above, for every $(s,t) \in \mathbb{T}\times \mathbb{T}$ pick a continuous, concave, and strictly increasing function $u_{st}:X\mapsto \mathbb{R}$ such that $u_{st}=u_{ts}$, $u_{st}(x^s)\geq u_{st}(y)$ for every $y \in B(p^s,w^s)$, and $u_{st}(x^t)\geq u_{st}(y)$ for every $y \in B(p^t,w^t)$.

Define a finite coalition structure $\Omega=\{V_{s}:s\in\mathbb{T}\}$, where $V_{s}\equiv\{u_{st}:t\in\mathbb{T}\}$ for every $s\in \mathbb{T}$. Then $\min_{u\in V_{s}}\left((u(x^{s})-u(y)\right)\geq 0$ for every $s\in \mathbb{T}$ and $y \in B(p^s,w^s)$. Thus, the CMU function $r^{\Omega}$ rationalizes $O^{T}$. Moreover, $\Omega$ is coherent because for any $s,t \in \mathbb{T}$, the function $u_{st}=u_{ts}$ belongs to $V_{s}\cap V_{t}$. Finally, continuity and strict monotonicity of $r^{\Omega}$ follow from Lemma \ref{Lem:Basic CMU}.

\subsubsection*{(iv) $\protect\Longrightarrow$ (v)}

Suppose that the dataset $O^{T}=\{(p^t,x^t)\}_{t\in \mathbb{T}}$ satisfies WGARP. Without loss of generality, assume that $T\geq 2$. Let $\mathcal {T}_2$ denote the collection of nonempty subsets of $\mathbb{T}$ that have $2$ elements. For every $\mathbb{S}\in \mathcal{T}_2$, WGARP restricted to the sample $O^2_\mathbb{S}$=$\{(p^s,x^s)\}_{s\in\mathbb{S}}$ implies GARP. Hence, by statement (iii) in Afriat's theorem, there exist numbers
$U_{\mathbb{S}}^{k}$ and $\lambda_{\mathbb{S}}^{k}>0$ for $k\in\mathbb{S}$ such
that $U_{\mathbb{S}}^{k}-U_{\mathbb{S}}^{l}\geq\lambda_{\mathbb{S}}^{k}p^{k}(x^{k}-x^{l})$ for every $k,l\in\mathbb{S}$.

For any ordered pair $(s,t)\in\mathbb{T}\times\mathbb{T}$ with $s\neq t$, set $R^{s,t}\equiv U_{\{s,t\}}^s-U_{\{s,t\}}^t$ and $\lambda^s_{s,t}\equiv \lambda^s_{\{s,t\}}$. Then we have $R^{s,t}=-R^{t,s}$ and $\lambda^s_{s,t}=\lambda^s_{t,s}$ because both of the ordered pairs $(s,t)$ and $(t,s)$ are associated with the set $\{s,t\}$. Also we have $R^{s,t}\geq\lambda_{s,t}^{s}p^{k}(x^{s}-x^{t})$ by construction. Finally, when $s=t$, we can simply let $R^{s,t}=0$ and $\lambda_{s,t}^{s}=1$.

\subsubsection*{(v) $\protect\Longrightarrow$ (vi)}

Suppose that statement (v) holds. Then $p^{s}(x^{s}-x^{t})\geq0$ implies $R^{s,t}\geq0$ because we have $\lambda_{s,t}^{s}>0$. Similarly $p^{s}(x^{s}-x^{t})>0$
implies $R^{s,t}>0$. Thus, we can let $V^{s,t}=R^{s,t}$ for every $s,t\in\mathbb{T}$.

\subsubsection*{(vi) $\protect\Longrightarrow$ (iv)}

Suppose that statement (vi) holds, but WGARP
fails, i.e., there exist some $s,t\in\mathbb{T}$ such that $p^{t}(x^{t}-x^{s})\geq0$ and $p^{s}(x^{s}-x^{t})>0$. Then the inequalities in statement (vi) imply $V^{t,s}\geq0$ and $V^{s,t}>0$, contradicting the condition
 $V^{s,t}=-V^{t,s}$ in statement (vi).

\subsection*{Proof of Lemma \ref{lem:CompensatedLD}}

Pick any $(p,w)\in P\times W$ and a bundle $x\in x_r(p,w)$. Given a new price vector $p' \in P$, set $w'\equiv p'x$. Since $r$ is strictly increasing and asymmetric, by Lemma  \ref{Lem:Asymmetry}, for any possible choice $x'\in x_r(p',w')$ the dataset $\{(p,x),(p',x')\}$ satisfies WGARP. By Claim \ref{budget interior}, the properties of $r$ also ensure that $p'x'=w'$, that is, $p'x'=p'x$. Then from WGARP we get $px'\geq px$, i.e., $0 \geq -p(x'-x)$. Summing this with the inequality $0\geq p'(x'-x)$ yields $0\geq (p'-p) (x'-x)$.

\subsection*{Proof of Lemma \ref{lem:No-WGARP}}

Throughout the proof, we write $r$ and $\succeq_r$ in place of $r^{\Omega'}$ and $\succeq_{\Omega'}$, respectively.

Every $u \in U$ induces a demand $x_u(p,w)=\left\{x\in B(p,w):u(x)\geq u(y)\; \forall y \in B(p,w)\right\}$ at $(p,w)\in P\times W$. Representation (\ref{justifiable}) implies  $\bigcup_{u\in U} x_u(p,w)\subseteq x_r(p,w)$ for every $(p,w)\in P\times W$. Moreover, for any $x,y\in X,$ we have $x\succ_r y$ iff $u(x)>u(y)$ for every $u \in U$. 

Let us first assume that there exist functions $u^1,u^2 \in U$ and a price-wealth pair $(p,w)\in P\times W$ s.t. 
\begin{equation}
\label{distinct ut}
x_{u^1}(p,w) \neq x_{u^2}(p,w).
\end{equation}
Since $x_r(\cdot,\cdot)$ contains both $x_{u^{1}}(\cdot,\cdot)$ and $x_{u^{2}}(\cdot,\cdot),$ statement (ii) will be proved if we can show that there exist price-wealth pairs $(p^1,w^1),(p^2,w^2) \in P\times W$ and bundles $x^1,x^2$ such that $x^1\in x_{u^{1}}(p^1,w^1), \; x^2\in x_{u^{2}}(p^2,w^2),\; p^1x^2<w^1,$ and $p^2x^1<w^2$.

Given the inequality (\ref{distinct ut}), without loss of generality, we may assume that the set $x_{u^1}(p,w) \setminus x_{u^2}(p,w)$ is nonempty. Pick a bundle $y^* \in x_{u^1}(p,w) \setminus x_{u^2}(p,w).$ Since $u^2$ is continuous and quasiconcave, the set $x_{u^2}(p,w)$ is compact and convex. Thus, by a separating hyperplane theorem, there exists a vector $q\in \mathbb{R}^{L}$ such that $qy^*<M\equiv \min\{qx:x\in x_{u^2}(p,w)\}.$ As $p\in P$ is a strictly positive vector, we may assume that $p+\varepsilon q$ also belongs to $P$ for every $\varepsilon \in (0,1].$ 

For every $n\in \mathbb{N},$ set $\tilde{p}^n = p+ \frac{1}{n}q$ and $\tilde{w}^n = w+ \frac{1}{2n}(M+qy^{*})$.  Note that $py^{*}=w$. Moreover, $qy^*<M$ implies $\frac{1}{n}qy^*<\frac{1}{2n}(M+qy^*)$. It follows that 
\[
\tilde{p}^n y^{*} = py^{*}+\frac{1}{n}qy^* < w+ \frac{1}{2n}(M+qy^*)=\tilde{w}^{n} \quad \text{for every } n\in \mathbb{N}.
\]

Select a bundle $z^n \in x_{u^2}(\tilde{p}^n,\tilde{w}^{n})$ for each $n \in \mathbb{N}.$ Note that $\lim_n (\tilde{p}^n,\tilde{w}^{n})=(p,w).$ Since $x_{u^{2}}(\cdot,\cdot)$ is a compact valued and upper hemicontinuous correspondence, the sequence $(z^n)$ has a subsequence $(z^{n_k})$ that converges to a bundle $\hat{x} \in x_{u^2}(p,w).$ Then $\lim_k qz^{n_k}=q\hat{x}\geq M > \frac{1}{2}(M+qy^*),$ which yields $qz^{n_k} > \frac{1}{2}(M+qy^*)$ for all sufficiently large $k \in \mathbb{N}.$ In what follows, we fix such a large index $k$.

By construction, we have 
\[
\tilde{p}^{n_k} z^{n_k} = pz^{n_k} + \frac{qz^{n_k}}{n_k} \leq \tilde{w}^{n_k}= w+ \frac{M+qy^{*}}{2n_k} \quad \text{and}\quad \frac{qz^{n_k}}{n_k} > \frac{M+qy^*}{2n_k}.
\]
These inequalities imply $pz^{n_k}<w$. Letting $(p^1,w^1)=(p,w),\; x^1= y^{*},\; (p^2,w^2)=(\tilde{p}^{n_k},\tilde{w}^{n_k}),$ and $x^2= z^{n_k}$ completes the proof of statement (ii).

Conversely, suppose now that property (\ref{distinct ut}) fails for every $u^1,u^2 \in U$ and $(p,w)\in P\times W$. Select an arbitrary function $\bar{u} \in U.$ Then, by hypothesis, $x_u(p,w)=x_{\bar{u}}(p,w)$ for every $u \in U$ and $(p,w)\in P\times W$. This implies 
\begin{equation}
\label{rational dem}
x_r(p,w)=x_{\bar{u}}(p,w) \quad \text{for every } (p,w)\in P\times W.
\end{equation}
Indeed, for any $y \in B(p,w) \setminus x_{\bar{u}}(p,w)$ and $x'\in x_{\bar{u}}(p,w)$, we have $x'\succ_r y$ because $u(x')>u(y)$ for every $u\in U$. This yields $x_{r}(p,w) \subseteq  x_{\bar{u}}(p,w)$. In turn, the converse inclusion $x_{\bar{u}}(p,w)\subseteq x_r(p,w)$ holds simply by definitions.

Finally, if (\ref{rational dem}) holds, then any dataset rationalized by $r$ is also rationalized by the standard utility function $\bar{u},$ which implies statement (i).

\subsection*{Proof of Lemma \ref{lem:No-WGARP2}}

Suppose every $u \in U$ is strictly quasiconcave, continuous and strictly increasing on $X$. Fix a pair $(p,w)\in P\times W$. Let $x^*_u$ denote the unique element of $x_u(p,w)$. For any $x\in B(p,w)$ that is distinct from $x^*_u$, we have $u(x)<u(x^*_u)$, and this implies $\neg(x\succeq_r x^*_u)$ by representation (\ref{incomplete MU}). It follows that $x^*_u\in x^+_r(p,w)$ for every $u\in U$. 

Since $(p,w)$ is an arbitrarily selected price-wealth pair, we conclude that $\bigcup_{u\in U} x_u(p,w)\subseteq x^+_r(p,w)$ for every $(p,w)\in P\times W$, which takes the role of the analogous inclusion that we have seen in the proof of Lemma \ref{lem:No-WGARP}. The remaining arguments in that proof can be repeated to prove Lemma \ref{lem:No-WGARP2} upon replacing the demand correspondence $x_r(p,w)$ with $x^+_r(p,w)$.

\subsection*{Proof of Theorem \ref{thm:Acyclicityk}}

Let us first prove a useful implication of Lemma \ref{Lem:Basic CMU}(ii), which is closely related to Claim~\ref{budget interior}.

\begin{claim}\label{cl:str_dec_y} Consider a CMU function $r=r^{\Omega}$, and suppose every coalition $U$ in $\Omega$ consists of strictly increasing functions. For any $(p,w)\in P\times W$, $x\in x_r(p,w)$ and $y\in X$ with $py<w$, we have $r(x,y)>0$.
\end{claim}

\begin{proof} If a bundle $y$ satisfies $py<w$, there exists a $z\in X$ such that $z\gg y$ and $pz<w$. The latter inequality implies $r(x,z)\geq 0$ for any $x\in x_r(p,w)$. Moreover, by Lemma~\ref{Lem:Basic CMU}(ii), if all coalitions in $\Omega$ consist of strictly increasing functions, then $r(x,\cdot)$ is a strictly decreasing function on $X$. It then follows that $r(x,y)> r(x,z)\geq 0.$
\end{proof}

We are now ready to prove Theorem \ref{thm:Acyclicityk}.

\subsubsection*{(i) $\protect\Longrightarrow$ (ii)}

Suppose that the dataset $O^{T}=\{(p^t,x^t)\}_{t\in \mathbb{T}}$ is rationalized by a finite CMU function $r^\Omega$, where $\Omega$ is a coalition structure with strictly increasing functions, and Nakamura number $\nu(\Omega)\geq k+1$.

Toward a contradiction, let us assume that the dataset $O^T$ violates $k$-acyclicity. Then $O^{T}$ contains a sequence of bundles $x^{t_1},x^{t_2},\dots,x^{t_k}$ such that $x^{t_1}\succeq^{R,D}x^{t_2}\succeq^{R,D}\cdots\succeq^{R,D}x^{t_k}$ and $x^{t_k}\succ^{R,D}x^{t_1}$. Our rationalization assumption implies $r^\Omega(x^{t_l},x^{t_{l+1}})\geq0$ for $l=1,\dots,k-1$. Then, by definition of $r^\Omega$, for each $l=1,\dots,k-1$, there exists a coalition $U_l\in \Omega$ such that $u(x^{t_l})\geq u(x^{t_{l+1}})$ for every $u \in U_l$. Moreover, by Claim \ref{cl:str_dec_y}, $x^{t_k}\succ^{R,D}x^{t_1}$ implies $r^\Omega(x^{t_k},x^{t_1})>0$. Hence, there exists a further coalition $U_k \in \Omega$ such that $u(x^{t_k})> u(x^{t_1})$ for every $u \in U_k$.  

Since $\nu(\Omega)\geq k+1$, any collection of $k$ coalitions in $\Omega$ has a nonempty intersection. Thus, there exists a function $u$ that belongs to $ \bigcap_{l=1}^{k}U_l$. But then we have $u(x^{t_1})\geq u(x^{t_2})\geq\cdots\geq u(x^{t_k})>u(x^{t_1})$, a contradiction.

\subsubsection*{(ii) $\protect\Longrightarrow$ (i)}

Suppose that the dataset $O^{T}=\{(p^t,x^t)\}_{t\in \mathbb{T}}$ satisfies $k$-acyclicity. If $T\leq k$, we can rationalize the dataset with a standard utility function, i.e., a degenerate coalition structure $\Omega=\{\{u\}\}$ with $\nu(\Omega)=\infty$. Thus, without loss of generality we may assume that $T>k$.

Let $\mathcal {T}_k$ denote the collection of nonempty subsets of $\mathbb{T}$ that have $k$ elements. Since $O^T$ satisfies $k$-acyclicity, for every $\mathbb{S}\in \mathcal{T}_k$ the sample $O_\mathbb{S}=\{(p^s,x^s)\}_{s\in\mathbb{S}}$ satisfies GARP. Hence, by Afriat's theorem, $O_\mathbb{S}$ is rationalized by a strictly increasing function $u_\mathbb{S}:X\mapsto \mathbb{R}$.  

For every $t\in\mathbb{T}$ define a set of functions $U_t\equiv\{u_\mathbb{S}:\mathbb{S}\in \mathcal{T}_k \text{ and } t\in \mathbb{S}\}$. Consider the coalition structure $\Omega\equiv \{U_t:t\in \mathbb{T}\}$. By construction, for any $\mathbb{S}\in \mathcal{T}_k$ and $t\in \mathbb{S}$, the bundle $x^t$ maximizes the function $u_\mathbb{S}$ on the budget $B(p^t,w^t)$. Since the coalition $U_t$ consists of such functions, it follows that the CMU function $r^\Omega$ rationalizes $O^T$.

It remains to show that $\nu(\Omega)\geq k+1$. Pick a collection of coalitions $\hat{\Omega}\subset \Omega$ with $|\hat{\Omega}|\leq k$, and set $l\equiv |\hat{\Omega}|.$ We claim that $\bigcap\{U: U\in \hat{\Omega}\}$ is nonempty. Without loss of generality, we may assume that $l=k$; otherwise we can add further coalitions into $\hat{\Omega}$. So, let $l=k$. Then, by definition of the coalitions in $\Omega$, there exist $k$ indices $t_1,\dots,t_k$ in $\mathbb{T}$ such that $\hat{\Omega}=\{U_{t_1},\dots,U_{t_k}\}$. Note that the set of indices $\mathbb{S}=\{t_1,\dots,t_k\}$ belongs to $\mathcal T_k$, and we have $t_j\in \mathbb{S}$ for $j=1,\dots,k$. This implies $u_\mathbb{S}\in U_{t_j}$ for $j=1,\dots,k$, as we wanted to show.

\subsection*{Proof of Theorem \ref{thm:Nak}}

Suppose that a finite dataset $O^{T}=\{(p^{t},x^{t})\}_{t\in \mathbb{T}}$ satisfies $L$-acyclicity. By Theorem \ref{thm:Acyclicityk}, $O^{T}$ is rationalized by a finite CMU function $r^{\Omega}$ with a Nakamura number $\nu(\Omega)\ \geq L+1.$

Fix a pair $(p,w)\in P\times W$. For any $U\in\Omega$ and $x\in X$, define a strictly preferred set
$P_{U}(x)=\{z\in X:u(z)>u(x)$ $\forall u\in U\}$. The set $P^{\Omega
}(x)=\bigcup\nolimits_{U\in\Omega}P_{U}(x)$ consists of all bundles that are
strictly preferred to $x$ by at least one coalition $U\in\Omega$. Note that we
have $z\in P^{\Omega}(x)$ if and only if $r^{\Omega}(z,x)>0.$

Set $B^{\ast}\equiv\{x\in X:px=w\}$. The dimension of the compact,
convex set $B^{\ast}$ equals $L-1$, which is less than or equal to $\nu(\Omega)-2$.
Moreover, each function $u$ in every coalition $U\in\Omega$ is continuous and
concave. Under these conditions, Theorem 1 of Schofield (1984) ensures that there exists an
$x^{\ast}\in B^{\ast}$ such that $B^{\ast}\cap P^{\Omega}(x^{\ast})=\emptyset$.

Let $(p^{T+1},x^{T+1})\equiv(p,x^{\ast})$. Toward a contradiction, suppose the extended dataset
$O^{T}\cup\{(p^{T+1},x^{T+1})\}$ violates WGARP. Then there exist indices
$s,t\in\{1,...,T+1\}$, such that
\[
s\neq t,\quad p^{s}x^{s}>p^{s}x^{t},\quad\text{and}\quad p^{t}x^{t}\geq
p^{t}x^{s}.
\]
Since $O^{T}$ is $L$-acyclic and $L\geq2$, we must have either $s=T+1$ or
$t=T+1$.

If $s=T+1$, we have $w^{T+1}>p^{T+1}x^{t}$, and there exists a $z\in B^{\ast}$ such that $z\gg
x^{t}$. Since $r^{\Omega}$ rationalizes $O^{T}$, the inequality $p^{t}%
x^{t}\geq p^{t}x^{s}$ implies $r^{\Omega}(x^{t},x^{s})\geq0$. As
$r^{\Omega}(x,x^{s})\ $is strictly increasing in $x$, it follows that
$r^{\Omega}(z,x^{s})>0$. Then $z$ belongs to $B^{\ast}\cap
P^{\Omega}(x^{s})$, contradicting the definition of $x^{\ast}%
=x^{s}.$

Finally, suppose $t=T+1$. By Claim \ref{cl:str_dec_y}, $p^{s}x^{s}>p^{s}x^{t}\ $implies
$r^{\Omega}(x^{s},x^{t})>0$. This means $r^{\Omega}(x^{s},x^{\ast})>0$.
Further, as $p^{T+1}x^{\ast}=p^{t}x^{t}\geq p^{t}x^{s}=p^{T+1}x^{s}$, there
exists a $z^{\prime}\in B^{\ast}$ such that $z^{\prime}\geq x^{s}$. Since $r(\cdot,x^{s})$ is strictly increasing, it follows that $r^{\Omega}(z^{\prime},x^{\ast})\geq r^{\Omega}%
(x^{s},x^{\ast})>0$. Then $z^{\prime}\in B^{\ast}\cap P^{\Omega}(x^{\ast})$,
which also contradicts the definition of $x^{\ast}$.

\subsection*{Proof of Lemma \ref{lem:expenditure}}
\subsubsection*{(i)}
Pick any $x\in X$ and suppose $x\notin h(p,x)$. In problem (\ref{Prb:E}), $x$ belongs to the constraint $\operatorname{cl} x^\uparrow$, so there must exist a $y\in \operatorname{cl} x^\uparrow$ such that $py<px$. As $y\in \operatorname{cl}x^\uparrow$, there also exists a $\hat{y} \in x^\uparrow$, sufficiently close to $y$, such that $p\hat{y}<px$. But then $\hat{y}$ is an element of $B(p,w_x)$ and $r(x,\hat{y})<0$, which means $x\notin x_r (p,w_x)$. 
 
 Conversely, consider now a bundle a $x$ such that $x\notin x_r(p,w_x)$. This means $r(x,y)<0$ for some $y\in X$ with $py\leq w_x$. Since $r(x,y)<0$, the assumptions on $r$ ensure that $y$ is non-zero. Let $y_\varepsilon\equiv (1-\varepsilon) y$ for $\varepsilon \in (0,1)$. The continuity of $r$ implies $r(x,y_\varepsilon)<0$ for all sufficiently small $\varepsilon$. Moreover, $py_\varepsilon=(1-\varepsilon)py<py$ because $p$ is strictly positive and the bundle $y$ is non-zero. It follows that for small $\varepsilon$, we have $y_\varepsilon \in x^\uparrow$ and $py_\varepsilon<w_x=px$, which imply $x\notin h(p,x)$.

\subsubsection*{(ii)} Problem (\ref{Prb:E}) has a solution by the Weierstrass theorem because the objective function $py$ is continuous, and the constraint set $\operatorname{cl}x^\uparrow $ can be replaced by the compact set $\{y\in \operatorname{cl}x^\uparrow:py\leq px \}.$

\subsubsection*{(iii)} Suppose $r(x,y)$ is strictly decreasing in $y$. Pick any $y\in X$ such that $r(x,y)\leq 0.$ Every neighborhood of $y$ contains a bundle $y'$ such that $y'\gg y$. Moreover, $r(x,y)\leq 0$ implies $r(x,y')<0$ and $y'\in x^\uparrow$. Hence, $y$ belongs to $\operatorname{cl}x^\uparrow$, and $\{y\in X: r(x,y)\leq 0\}$ is a subset of $\operatorname{cl}x^\uparrow$.

Conversely, pick any $y\in \operatorname{cl}x^\uparrow$. Then there exists a sequence $\{y^n\}$ in $x^\uparrow$ that converges to $y$. Since $y^n$ belongs to $ x^\uparrow$, we have $r(x,y^n)<0$ for every $n\in \mathbb{N}$. Then $r(x,y)=\lim_n r(x,y^n)\leq 0$, by the continuity of $r$. Hence, $\{y\in X: r(x,y)\leq 0\}$ contains $\operatorname{cl}x^\uparrow$.

Now, pick any $y\in h(p,x)$. We claim that $r(x,y)=0$. Otherwise, $r(x,y)< 0$ by the first part of the proof. This implies $y\neq 0$, for we have $r(x,0)\geq 0$. Since $r$ is continuous, $r(x,y)< 0$ also implies $r(x,\gamma y)<0$ for a $\gamma\in (0,1)$ sufficiently close to $1$. Then $\gamma y$ belongs to $x^\uparrow $ and $p(\gamma y)<py$, which contradict the hypothesis $y\in h(p,x)$.     

\bigskip

\bibliographystyle{ecta}
\bibliography{WARPpaper}

\end{document}

\title{Online Appendix to \\
 ``A Rationalization of the Weak Axiom of Revealed Preference''\\
 ONLINE APPENDIX}
\author{Victor H. Aguiar\thanks{Department of Economics, Simon Fraser University, British Columbia, Canada, vaguiar@sfu.ca.} \and Per Hjertstrand\thanks{Research Institute of Industrial Economics (IFN), Stockholm, Sweden,
per.hjertstrand@ifn.se. } \and Roberto Serrano\thanks{Department of Economics, Brown University, Providence, Rhode Island,
U.S.A., roberto\_serrano@brown.edu.} \and Ozgur Evren\thanks{New Economic School, Moscow, Russian Federation, oevren@nes.ru.} }
\date{This version: November 2025}

\maketitle

\appendix

%\section{Another Example to Illustrate the CMU Model}

%We provide another example that connects our work with the recent models of \citet{hara2019coalitional}
%and \citet{frick2019boolean}. Consider a case where there is one physical good (money) and $3$
%states of the world (good, business-as-usual, bad). The Arrow-Debreu
%state-contingent securities are $x_{l}$, for $l=1,2,3$. Suppose
%that the consumer knows that state $1$ happens with probability $1/2$,
%and state $2$ or $3$ happens with probability $1/2$ but there is
%uncertainty about which of these two will occur. The consumer has
%two moods, $U_{r}=\{u_{1,r},u_{2,r}\}$ and $U_{m}=\{u_{1,m},u_{2,m}\}$,
%where $r$ stands for realistic and $m$ for pessimistic. The utility
%of an Arrow-Debreu security $x$ associated with the different utilities
%$i\in\{1,2\}$ and $j\in\{r,m\}$ is:
%\[
%u_{i,j}=1/2v(x_{1})+(1/2-\pi_{i,j})v(x_{2})+\pi_{i,j}v(x_{3})=E_{\pi_{i,j}}[v(x)],
%\]
%where $0\leq\pi_{i,j}\leq1/2$ and $v$ is a Bernoulli utility defined
%over money. The subjective probabilities are: $\pi_{1,r}=1/5$, $\pi_{1,m}=1/3$,
%and $\pi_{2,r}=\pi_{2,m}=1/4$. We note that this model is similar
%to the ambiguity framework posed in \citet{frick2019boolean}, where
%the ``act'' corresponds to choosing bundle $x$ over bundle $y$,
%such that the utility of such ``act'' is given by:
%\[
%r(x,y)=\max_{j\in\{r,m\}}\min_{i\in\{1,2\}}(E_{\pi_{i,j}}(v(x)-v(y))).
%\]

\section{WARP and Single-Valued Demand Functions}
\label{WARPCharac}

Consider a dataset $O^{T}=\{(p^{t},x^{t})\}_{t\in\mathbb{T}}$. \citeauthor{Houthakker1950}'s (\citeyear{Houthakker1950}) strong axiom is the analogue of GARP for single-valued demand functions:

\begin{axiom}
(SARP)
The strong axiom of revealed preference (SARP)
holds if there is no pair of observations $s,t\in\mathbb{T}$ such
that $x^{t}\neq x^{s}$, $x^{t}\succeq^{R}x^{s}$ and $x^{s}\succeq^{R,D}x^{t}$.
\end{axiom}

The binary version of SARP is known as the weak axiom:

\begin{axiom}
(WARP)
The weak axiom of revealed preference (WARP)
holds if there is no pair of observations $s,t\in\mathbb{T}$ such
that $x^{t}\neq x^{s}$, $x^{t}\succeq^{R,D}x^{s}$ and $x^{s}\succeq^{R,D}x^{t}$.
\end{axiom}

\citet{matzkin_testing_1991-1} shows that a dataset that satisfies SARP can be rationalized by a strictly concave utility function. In this appendix we provide a characterization of WARP, along the lines of Theorem 1, by utilizing \citeauthor{matzkin_testing_1991-1}'s (\citeyear{matzkin_testing_1991-1}) theorem in place of Afriat's theorem. The corresponding CMU function involves strictly concave functions and satisfies a form of local strict concavity:

\begin{definition}\label{Def1} We say that a preference function $r:X\times X \mapsto \mathbb{R}$ is \emph{strictly star-concave} if $r(\alpha x+(1-\alpha)y,x)>\alpha r(x,x)+(1-\alpha)r(y,x)$ for every $\alpha \in (0,1)$, and $x,y \in X$ with $x\neq y$.\footnote{A function $f:X\mapsto \mathbb{R}$ is said to be strictly star-concave around a point $x\in X$ if $f(\alpha x+(1-\alpha)y)>\alpha f(x)+(1-\alpha)f(y)$ for every $\alpha \in (0,1)$, and $x,y \in X$ with $x\neq y$. In Definition \ref{Def1}, we demand this property from $f\equiv r(\cdot,x)$ exactly at the point $x$, as specified by the second argument of $r$.} \end{definition}

The following observation takes the roles of Lemmas 1 and 2 of the main text.

\begin{customlem}{A1}\label{LemA1}
\begin{enumerate}
\item A CMU function $r^\Omega$ is strictly star-concave provided that every coalition $U \in \Omega$ consists of strictly concave functions.
\item If a dataset $O^T$ is rationalizable by a strictly star-concave and asymmetric
preference function $r$, then $O^T$ satisfies WARP.    
\end{enumerate}
\end{customlem}

\begin{proof}

\noindent \emph{(i)} Consider a compact coalition structure $\Omega$, and suppose every $U\in \Omega$ consists of strictly concave functions. For any $x,y\in X$, let $r_U(x,y)=\min_{u\in U}(u(x)-u(y))$, so that $r^\Omega (x,y)=\max_{U\in \Omega} r_{U}(x,y)$.

Fix any $\alpha \in (0,1)$ and $x,y \in X$ with $x\neq y$. Since $r^\Omega (x,x)=0$, strict star-concavity is equivalent to the condition $r^\Omega(z,x)>(1-\alpha)r^\Omega(y,x)$, where $z\equiv \alpha x+(1-\alpha)y$.

Let $\hat{U} \in \Omega$ and $\hat{u}\in U$ be such that $r^\Omega (y,x)=r_{\hat{U}}(y,x)$, and $r_{\hat{U}}(y,x)=\hat{u}(y)-\hat{u}(x).$ Strict concavity of the functions in $\hat{U}$ implies $u(z)-u(x)>(1-\alpha)(u(y)-u(x))$ for every $u\in \hat{U}.$ Since $u(y)-u(x) \geq \hat{u}(y)-\hat{u}(x)$ for $u\in \hat{U}$, it follows that $r^\Omega(z,x)\geq \min_{u\in \hat{U}}(u(z)-u(x))>(1-\alpha)(\hat{u}(y)-\hat{u}(x))=(1-\alpha)r^\Omega(y,x)$, as we wanted to show.\medskip

\emph{(ii)} Let $r$ be a strictly star-concave and asymmetric preference function. Pick any $(p,w)\in P\times W$ and $x\in x_r(p,w)$. Then $r(x,y) \geq 0$ for every $y\in B(p,w)$. In particular, this yields $r(x,x)\geq 0$. We claim that $x\succ_r y$ for every $y\in B(p,w)$ with $y\neq x$. Suppose by contradiction that $r(y,x)\geq 0$ for some $y\in B(p,w)\setminus\{x\}$. Let $z=\alpha x+(1-\alpha)y$, and note that we have $r(z,x)>\alpha r(x,x) + (1-\alpha) r(y,x)\geq 0$ by strict star-concavity. Then from asymmetry, we also get $r(x,z)<0$, which contradicts the optimality condition $x\in x_r(p,w).$

Now, suppose that the function $r$ rationalizes a dataset $O^T$ that violates WARP. Let $s,t\in\mathbb{T}$ be such
that $x^{t}\neq x^{s}$, $x^{s}\in B(p^t,w^t)$, and $x^{t}\in B(p^s,w^s)$, By the first part of the proof, the rationalization condition $x^{s}\in x_r(p^s,w^s)$ implies $x^s\succ_r x^t$, and this contradicts the optimality of $x^t$ in $B(p^t,w^t)$.

\end{proof}

\begin{remark}  Consider a strictly star-concave preference function $r$, and a pair of bundles $x,y$ such that $x\sim_r y$. Assume further that $y\succeq_r y$, that is, $r(y,y)\geq 0$. Then any neighborhood of $x$ contains a bundle $x'=\alpha x +(1-\alpha)y$ for some $\alpha \in (0,1)$ such that $r(x',y)>0$. This local non-satiation condition is implicit in statement (ii) of Lemma \ref{LemA1}, and it takes the role played by the strict monotonicity property in Lemma 2 of the main text.
\end{remark}

The next theorem characterizes WARP along the lines of Theorem 1 in the main text.

\begin{customthm}{A1}\label{ThmA1}
Consider a finite dataset $O^{T}=\{(p^{t},x^{t})\}_{t\in\mathbb{T}}$. The following
statements are equivalent:
\begin{enumerate}
\item 
The dataset $O^{T}$ can be rationalized by an asymmetric, strictly star-concave, and continuous preference function $r:X\times X \mapsto \mathbb{R}$.
\item The dataset $O^{T}$ can be rationalized by a coherent, strictly star-concave,  and continuous CMU function.
\item The dataset $O^{T}$ can be rationalized by a skew-symmetric, coherent, and continuous CMU function $r^{\Omega}$ such that each coalition in $\Omega$ consists of strictly increasing and strictly concave functions.
\item The dataset $O^{T}$ satisfies WARP.
\end{enumerate}
\end{customthm}
     
    The implications (iii) $\Longrightarrow$ (ii) $\Longrightarrow$ (i) $\Longrightarrow$ (iv) are direct consequences of Lemma \ref{LemA1} above and Lemma 3 in the main text. In turn, that (iv) implies (iii) can be established as in the proof of Theorem 1. Here, the only novelty is the strict concavity of the functions $u_{st}:X\mapsto \mathbb{R}$ that appear in the utility matrix $\mathcal{M}=[u_{st}]$ and the rationalizing coalition structure $\Omega(\mathcal{M})$. These functions are obtained by applying \citeauthor{matzkin_testing_1991-1}'s (\citeyear{matzkin_testing_1991-1}) theorem to binary subsets of the dataset $O^T$.     

\subsection{Discussion}
    If a preference function $r:X\times X \mapsto \mathbb{R}$ is asymmetric, strictly star-concave, and also reflexive, then the associated preference relation $\succeq_r$ is \emph{strictly star-convex} in the following sense: 
    \[y\succeq_r x \text{ and } x\neq y\quad \Longrightarrow \quad \alpha x+(1-\alpha)y\succ_r x\; \forall x,y\in X \text{ and } \alpha \in (0,1).\]
    Following the proof of Lemma \ref{LemA1}, it is easily checked that rationalization by such a preference relation $\succeq$, i.e., the condition $x^t\succeq y$ for every $y\in B(p^t,w^t)$ and $t\in \mathbb{T}$, implies WARP. Thus, from Theorem \ref{ThmA1} it also follows that WARP is fully characterized with rationalization by a strictly star-convex preference relation.
    
    Theorem 16 of \citet{kim_nontransitive-nontotal_1986-1} shows that even for an infinite dataset, WARP implies a rationalization by a strictly star-convex preference relation. The preference relation constructed in their proof is piecewisely defined, and partly includes the lexicographic relation on $\mathbb{R}^L$. By contrast, our proof technique, which we also used in Theorem~1, provides a continuous preference relation represented by a well-behaved preference function.

    Finally, it should be noted that the strict star-convexity of a preference relation is quite distinct from the $>$-consistency property that underlies our characterization of WGARP in Theorem 1. In the remainder of this appendix, we will study this consistency property in a more general framework.

\section{Abstract Choice Settings}
\label{subsec:abstract}

\citet{nishimura2017comprehensive} have extended Afriat's theorem to a general framework that allows one to study rationalizability in alternative choice problems, such as a choice of lotteries or time-contingent consumption. In this appendix, we provide an analogous extension of Theorem 1. This setup also covers nonlinear budget sets as in \cite{Forges2009}, and discrete choice sets consisting of commodity bundles with integer coordinates.

Consider a locally compact and separable metric space $X$, equipped with a partial order $\unrhd$ (a reflexive, transitive and antisymmetric binary relation) that is closed as a subset of $X\times X$. The relation $\unrhd$ represents a dominance criterion that must be respected by every rational individual. In classical consumer theory, we have $X=\mathbb{R}^L_+$ and $\unrhd$ is the coordinate-wise order $\geq$. The current setting also allows for different specifications.\footnote{ For example, $X$ can be the space of all countably additive, Borel probability meaures on $[0,1]$ equipped with the weak-convergence metric. In this case, we can let $\unrhd$ be the first-order stochastic dominance relation. Or, $X$ can be a set $A\times D$, where $a\in A$ and $\tau \in D$ represent the amount and date of consumption, respectively. One may then define a partial order $\unrhd$ on $X$ by the rule $(a,\tau)\unrhd (a',\tau')$ if and only if $a\geq a'$ and $\tau \leq \tau'$.}

We denote by $\vartriangleright$ the strict part of $\unrhd$, that is, for $x,z \in X$ we have $x\vartriangleright z$ if and only if $x\unrhd z $ and $x\neq z$. A function $u:X \mapsto \mathbb{R}$ is strictly $\vartriangleright$-increasing (respectively, strictly $\vartriangleright$-decreasing) if $x\vartriangleright z$ implies $u(x)>u(z)$ (respectively, $u(x)<u(z)$). We say that a preference function $r:X\times X$ is \emph{strictly $\vartriangleright$-monotonic} if $r(\cdot,y)$ is strictly $\vartriangleright$-increasing and $r(y,\cdot)$ is strictly $\vartriangleright$-decreasing for every $y\in X$. (We clarify below why this two-sided monotonicity condition is useful.)

%Similarly, a preference function $r:X\times X \mapsto \mathbb{R}$ is \emph{strictly $\unrhd$-increasing} if $r(\cdot,y)>r(\cdot,y)$ is strictly $\unrhd$-increasing for every $y \in X$. 

The asymmetry and skew-symmetry of a preference function $r$, as well as the associated preference relation $\succeq_r$ are defined as in the main text. Given a compact set $A\subseteq X$ of feasible alternatives and a preference function $r$, the set of optimal alternatives is $x_r(A)\equiv\{x\in A:r(x,y)\geq0,\; \forall\ y\in A\}$.

We also define a CMU function $r^\Omega$ as in the main text, for a nonempty compact (relative to Hausdorff metric) collection $\Omega$ of nonempty compact sets $U\subset C(X)$. Here, $C(X)$ is the space of real valued, continuous functions on $X$ equipped with a metric that induces the compact convergence topology.

A \emph{dataset} $O^T$ refers to a finite collection of ordered pairs $(A^t,C^t)$ for $t\in \mathbb{T}\equiv \{1,...,T\}$, where $A^t\subseteq X$ is a compact set of feasible alternatives, and $C^t\subseteq A^t$ is a nonempty, compact set of alternatives chosen from the menu $A^t$ in the observation $t\in \mathbb{T}.$ 

Since the menus are not necessarily comprehensive, we need to modify revealed preference definitions. Pick any $x^t\in C^t$ and $x^s\in C^s$ for some $s,t\in \mathbb{T}$. We say that
$x^t$ is \emph{directly revealed preferred to} $x^{s}$, written $x^{t}\succeq^{R,D}x^{s}$,
when there exists a $y^t\in A^t$ such that $y^t \unrhd x^{s}$. In turn, $x^{t}$ is \emph{strictly and
directly revealed preferred to} $x^{s}$, written $x^{t}\succ^{R,D}x^{s}$,
when there exists a $y^t\in A^t$ such that $y^t \vartriangleright x^{s}$.\footnote{It is readily checked that these definitions of direct revealed preference coincide with their counterparts in the main text when $X=\mathbb{R}^L_+$, the partial order $\unrhd$ is the coordinate-wise order $\geq$, the menu $A^t$ equals the budget set $B(p^t,p^tx^t)$ for a strictly positive price vector $p^t$, and $C^t=\{x^t\}$ for every $t\in \mathbb{T}$.}

The following axiom is the analogue of GARP introduced by \citet{nishimura2017comprehensive}. The novelty of the axiom is the role of the dominance relation $\unrhd$ in direct revealed preference relations defined above. 
\begin{axiom}
 (Cyclical $\unrhd$-Consistency)
  The \emph{cyclical $\unrhd$-consistency axiom} holds if the dataset $O^T$ does not contain a chain $(x^{t_1},\ldots,x^{t_n})\in C^{t_1}\times \cdots\times C^{t_n}$ such that $x^{t_1}\succeq^{R,D}x^{t_2}\succeq^{R,D}\cdots\succeq^{R,D}x^{t_n}$ and $x^{t_n}\succ^{R,D}x^{t_1}$.
\end{axiom}

In this context, WGARP corresponds to a binary version of cyclical $\unrhd$-consistency:  

\begin{axiom}
 (Weak Cyclical $\unrhd$-Consistency)
  The \emph{weak cyclical $\unrhd$-consistency axiom} holds if there is no pair of indices $s,t\in\mathbb{T}$ and bundles $(x^{s},x^{t})\in C^{s}\times C^{t}$ 
such that $x^{t}\succeq^{R,D}x^{s}$ and $x^{s}\succ^{R,D}x^{t}$.
\end{axiom}

Theorem 2 of \citet{nishimura2017comprehensive} shows that cyclical $\unrhd$-consistency axiom\footnote{To be precise, cyclical $\unrhd$-consistency axiom, as stated above, reformulates \citeauthor{nishimura2017comprehensive}'s (\citeyear{nishimura2017comprehensive}) axiom in the GARP format; the relation between the two axiom statements is analogous to that between \citeauthor{afriat_construction_1967}'s (\citeyear{afriat_construction_1967}) original axiom and \citeauthor{varian_nonparametric_1982}'s (\citeyear{varian_nonparametric_1982}) GARP. As a more significant difference, here we focus on a partial order $\unrhd$, while \citeauthor{nishimura2017comprehensive} also allow for ties in the dominance relation.} is equivalent to rationalizability by a strictly $\vartriangleright$-increasing and continuous function $u:X\mapsto \mathbb{R}$ in the following sense:

\begin{definition}
  (Utility rationalization)
   \label{UtilityRat}
   A function $u:X\mapsto\mathbb{R}$ rationalizes a dataset $O^{T}=\{(A^{t},C^{t})\}_{t\in\mathbb{T}}$ if $u(x^{t})\geq u(y)$ for every $y\in A^t$, $x^t\in C^t$, and $t\in\mathbb{T}$.
\end{definition}
This definition is adapted to preference functions as follows:
\begin{definition}
  (Preference function rationalization)
   \label{PrefFuncRat}
   A preference function $r:X\times X\mapsto\mathbb{R}$ rationalizes a dataset $O^{T}=\{(A^{t},C^{t})\}_{t\in\mathbb{T}}$ if $C^t\subseteq x_r(A^t)$ for every $t\in\mathbb{T}$.
\end{definition}

The changes in the definitions of direct revealed preference relations necessitate the following modification of Lemma 2, which demands a two-sided monotonicity condition. 

\begin{customlem}{A2}
\label{Lem:Asymmetry}
If a dataset $O^T$ is rationalizable by a strictly $\vartriangleright$-monotonic and asymmetric preference function $r$, then $O^T$ satisfies the weak cyclical $\unrhd$-consistency axiom.
\end{customlem}

\begin{proof}
Consider a strictly $\vartriangleright$-monotonic and asymmetric preference function $r$. 
Let $x,y,z\in X$ be such that $x\succeq_r z$ and $z\vartriangleright y$. The former property implies $r(x,z)\geq 0$, and the latter implies $r(x,y)>r(x,z)$ because $r(x,\cdot)$ is strictly $\vartriangleright$-decreasing. It follows that $r(x,y)>0$.

Similarly, if $x\vartriangleright z$ and $z\succeq_r y$, then we have $r(x,y)>0$ because $r(\cdot,y)$ is strictly $\vartriangleright$-increasing.

Suppose now that the preference function $r$ rationalizes a dataset $O^T$. Pick a pair of data points $x^t\in C^t$ and $x^s\in C^s$ such that $x^t\succeq^{R,D} x^s$. Then there exists a $z\in A^t$ such that $z\unrhd x^s$. If $z=x^s$, we have $r(x^t,x^s)\geq 0$ because $x^t$ belongs to $x_r(A^t)$ by rationalization. In turn, if $z\neq x^s$, we have $z\vartriangleright x^s$ and $x^t\succ^{R,D} x^s$. In this case, the properties $x^t\succeq_r z$ and $z\vartriangleright x^s$ jointly imply $r(x^t,x^s)>0$, as we have seen in the first part of the proof.

Symmetrically, $x^s\succ^{R,D} x^t$ implies $r(x^s,x^t)>0$, and also $r(x^t,x^s)<0$ by asymmetry of the function $r$.

It follows that $x^t\succeq^{R,D} x^s$ and $x^s\succ^{R,D} x^t$ jointly imply $r(x^t,x^s)\geq 0$ and $r(x^t,x^s)<0$, which cannot hold simultaneously.
\end{proof}

It should also be noted that the arguments used in the proofs of Lemmas 1, 3 and 4 of the main text remain true on any metric space $X$ equipped with a partial order $\unrhd$.\footnote{The only exception is the fact that the compact convergence topology on $C(X)$ may not be metrizable unless $X$ is locally compact and separable, as we assume in this appendix. These topological conditions are also demanded by Theorem 2 in \citet{nishimura2017comprehensive}.} We are now ready to state characterizations of the weak cyclical $\unrhd$-consistency axiom.  

\begin{customthm}{A2}\label{ThmA2}
Consider a dataset $O^{T}=\{(A^{t},C^{t})\}_{t\in\mathbb{T}}$, and suppose that the alternatives belong to a locally compact and separable metric space $X$, equipped with a closed partial order $\unrhd$. Then the following
statements are equivalent:
\begin{enumerate}
\item 
The dataset $O^{T}$ can be rationalized by an asymmetric, strictly $\vartriangleright$-monotonic, and continuous preference function $r:X\times X \mapsto \mathbb{R}$.
\item The dataset $O^{T}$ can be rationalized by a coherent, strictly $\vartriangleright$-monotonic,  and continuous CMU function.
\item The dataset $O^{T}$ can be rationalized by a skew-symmetric, coherent, and continuous CMU function $r^{\Omega}$ such that each coalition in $\Omega$ consists of strictly $\vartriangleright$-increasing functions.
\item The dataset $O^{T}$ satisfies the weak cyclical $\unrhd$-consistency axiom.
\end{enumerate}
\end{customthm}

By Theorem 2 of \citet{nishimura2017comprehensive}, the weak cyclical $\unrhd$-consistency axiom applied to a binary sample $\{(A^k,C^k)\}_{k=s,t}$ allows us to rationalize this sample by a function $u_{st}$ in the sense of Definition \ref{UtilityRat}. This means that every $x^k\in C^k$ maximizes $u_{st}$ on the set $A^k$ for $k=s,t$. Then, holding $s$ fixed, every $x^s\in C^s$ maximizes on the set $A^s$ any function of the form $\sum_{t\in\mathbb{T}}\mu_t u_{st}$ with nonnegative weights $\mu_t$. This allows one to derive statement (iii) from statement (iv), just as in Theorem 1 of the main text. The remaining implications in Theorem \ref{ThmA2} are direct consequences of Lemmas 1, 3 and 4 in the main text, and Lemma \ref{Lem:Asymmetry} above.

\section{Ordinal Content of Coherent and Strictly Monotonic CMU Functions}
\label{subsec:ordinal}
%% Representation
Let $X$ be a locally compact and separable metric space equipped with a closed partial order $\unrhd$. In Theorem \ref{ThmA2}, we have seen that the weak cyclical $\unrhd$-consistency axiom is equivalent to rationalizability by a coherent CMU function with coalitions that consist of strictly $\vartriangleright$-increasing functions. In this appendix, we analyze the structure of preference relations on $X$ that can be represented by such preference functions in a general form that does not require a compact coalition structure.

In what follows, $\succeq$ denotes a reflexive binary relation on $X$ that represents the preferences of a decision maker. We say that $\succeq$ admits a \emph{generalized CMU representation} if there exists a nonempty collection $\Omega$ of nonempty sets $U\subseteq C(X)$ such that for every  $x,y \in X$, 

\begin{equation}
x\succeq y\quad\Longleftrightarrow\quad\exists U\in\Omega:\;u(x)\geq u(y)\;\forall u\in U.\label{1}%
\end{equation}

The dominance relation $\unrhd$ and the preference relation $\succeq$ jointly define a
binary relation $\succ^{\ast}$ on $X$ by the following rule: For any $x,y\in X,$%
\[
x\succ^{\ast}y\quad\Longleftrightarrow\quad\exists z\in X\; \text{such that}\;x\vartriangleright z\succeq y,\;\text{or,}\;x\succeq z\vartriangleright y\text{.}
\]
Throughout the paper, we have seen an interplay between asymmetry of a preference function $r$, its strict monotonicity with respect to the dominance relation $\vartriangleright$, the consistency property $x\succ^{*} y \Longrightarrow r(x,y)>0$, and a further consistency property $x\succ^{*} y \Longrightarrow x\succ_r y$. Here, we focus on related properties of the order pair $(\succeq,\succ^*)$.
\begin{definition} 
   \begin{enumerate}   
   \item $(\succeq,\succ^*)$ is \emph{weakly consistent} if $x\succ^{*}y$ implies $x\succeq y$ for every $x,y\in X$.
   \item $(\succeq,\succ^*)$ is \emph{weakly acyclic} if there is no $x,y\in X$ such that  $x\succ^* y$ and $y\succeq x$. 
   \item $(\succeq,\succ^*)$ is \emph{strictly consistent} if it is weakly consistent and weakly acyclic.
   \end{enumerate}
\end{definition}

Property (ii) above is a binary version of an acyclicity condition for order pairs studied by \citet{Chambers2016}. Properties (i) and (ii) jointly mean that $x\succ^{*}y$ implies $x\succ y$, which is our strict consistency condition.

\begin{customthm}{A3}
\label{ThmRep}
Let $X$ be a locally compact and separable metric space, equipped with a closed partial order $\unrhd$. The following statements are equivalent for a binary relation $\succeq$ on $X$:
\begin{enumerate}
\item $\succeq$ is reflexive and the order pair $(\succeq,\succ^*)$ is strictly consistent.
\item $\succeq$ admits a generalized CMU representation with a coherent coalition structure $\Omega$ such that each coalition in $\Omega$ consists of strictly $\vartriangleright$-increasing functions. 
\end{enumerate}
\end{customthm}

Together, Theorems \ref{ThmA2} and \ref{ThmRep} show that the weak cyclical $\unrhd$-consistency axiom makes the observed choices compatible with a preference relation $\succeq$ such that the order pair $(\succeq,\succ^*)$ is strictly consistent.

We will derive Theorem \ref{ThmRep} from \citeauthor{evren2021extension}'s (\citeyear{evren2021extension}) characterization of generalized CMU representations that consist of strictly increasing component functions. The novelty of the present exercise is the coherence condition in statement (ii) of Theorem~\ref{ThmRep}.

\subsection*{Proof of Theorem \ref{ThmRep}}
We first prove a claim concerned with strictly $\vartriangleright$-increasing functions on a set $Z\subseteq X$. As usual, we say that a function $f:Z\mapsto\mathbb{R}$ is
strictly $\vartriangleright$-increasing if $z\vartriangleright z'$ implies $f(z)>f(z')$ for every $z,z'\in Z$.

\begin{customclaim}{A1}\label{claim:extension}
 Suppose the relation $\succeq$ is reflexive and
the pair $(\succeq,\succ^*)$ is strictly consistent. Then for any $x,y,x^{\prime},y^{\prime}\in X$
with $x\succeq y$ and $x^{\prime}\succeq y^{\prime}$, there exists a strictly
$\vartriangleright$-increasing function $f:\{x,y,x^{\prime},y^{\prime}\}\mapsto\mathbb{R}%
$ such that $f(x)\geq f(y)$ and $f(x^{\prime})\geq f(y^{\prime
})$.
\end{customclaim}

\begin{proof}
From reflexivity of $\succeq$ and strict consistency of $(\succeq,\succ^*)$, it easily follows that the relation $\succeq$ is strictly $\unrhd$-monotonic in the following sense: For any $x,y \in X$ with $x\vartriangleright y$ (respectively, $x\unrhd y$) we have $x\succ y$ (respectively, $x\succeq y$).

Pick any
$x,y,x^{\prime},y^{\prime}\in X$ with $x\succeq y$ and $x^{\prime} \succeq y^{\prime}$. Let
$Z=\{x,y,x^{\prime},y^{\prime}\}$. If $x^{\prime} \succ y^{\prime}$, then strict $\unrhd$-monotonicity of $\succeq$, and weak consistency of $(\succeq,\succ^*)$ allow us to invoke Claim~$2$ of
 \citet{evren2021extension}. This claim yields a strictly $\vartriangleright$-increasing function $f:Z\rightarrow
\mathbb{R}$ such that $f(x)\geq f(y)$ and $f(x^{\prime})>f(y^{\prime})$. Hence, without loss of generality we may assume $x \sim y$ and $x^{\prime} \sim y^{\prime}$.
Then we cannot have $x\vartriangleright y$, $y \vartriangleright x$, $x^{\prime} \vartriangleright y^{\prime}$ or
$y^{\prime} \vartriangleright x^{\prime}$ because $\succeq$ is strictly $\unrhd$-monotonic.

Consider the constant function $\mathbf{1}_{Z}$. If this function is strictly
$\vartriangleright$-increasing, we are done. Assume therefore that $\mathbf{1}_{Z}$ is not
strictly $\vartriangleright$-increasing. This means there exist $z,z^{\prime}\in Z$ such that
$z \vartriangleright z^{\prime}$. Then either we have $z\in\{x,y\}$ and $z^{\prime}%
\in\{x^{\prime},y^{\prime}\}$, or else, $z^{\prime}\in\{x,y\}$ and
$z\in\{x^{\prime},y^{\prime}\}$. Without loss of generality, let us assume
$z\in\{x,y\}$ and $z^{\prime}\in\{x^{\prime},y^{\prime}\}$.

Note that if there exists a $w\in\{x,y\}\cap\{x^{\prime},y^{\prime}\}$, we
must have $w \sim z$ and $w \sim z^{\prime}$. But since the pair $(\succeq ,\succ^{*})$ is strictly consistent, $w \sim z \vartriangleright z^{\prime}$ implies $w \succ z^{\prime}$, which contradicts the condition
$w \sim z^{\prime}$. Thus, we conclude that the sets $\{x,y\}$ and $\{x^{\prime
},y^{\prime}\}$ are disjoint.

We can now define a function $f:Z\rightarrow\mathbb{R}$ as $f(w)=1$ for
$w\in\{x,y\}$, and $f(w)=0$ for $w\in\{x^{\prime},y^{\prime}\}$. The final
step is to show that $f$ is strictly $\vartriangleright$-increasing. Otherwise, there exist
$w,w^{\prime}\in Z$ such that $w^{\prime}\vartriangleright w$ and $f(w)\geq f(w^{\prime})$.
Clearly, these two properties jointly imply $w\in\{x,y\}$ and $w^{\prime}%
\in\{x^{\prime},y^{\prime}\}$. Since $(\succeq ,\succ^{*})$ is strictly consistent, from
$w^{\prime}\vartriangleright w \sim z$ we get $w^{\prime} \succ z$, while $z \vartriangleright z^{\prime}\sim w^{\prime}$
implies $z\succ w^{\prime}$, a contradiction. Hence, $f$ is strictly $\vartriangleright$-increasing, as
we claimed.
\end{proof}
Now we are ready to prove Theorem~\ref{ThmRep}.
\par

%%%%%%%%%%%%%%%%%%%%New Proof
\subsubsection*{(i) $\implies$ (ii)}
\par
For every $x,y\in X$ with $x\succeq y$, let $U_{(x,y)}$ denote the
collection of all continuous, strictly $\vartriangleright$-increasing functions $u:X\mapsto
\mathbb{R}$ such that $u(x)\geq u(y)$. Define $\Omega=\{U_{(x,y)}:x\succeq y\}$.

By assumptions, $X$ is a locally compact, separable (and second countable) metric space, while $\unrhd$ is a closed partial on $X$. Thus, by Corollary~$3$ in \citet{evren2021extension}, any strictly
$\vartriangleright$-increasing function $f$ on a finite set $Z\subseteq X$ can be extended to a
strictly $\vartriangleright$-increasing and continuous function $u$ on $X$.

This extension result and Claim \ref{claim:extension} jointly imply that
any pair of sets $U_{(x,y)}$ and $U_{(x',y')}$ in $\Omega$ contain a common element $u$. Also, by reflexivity of $\succeq$, $\Omega$ is a nonempty collection of nonempty sets. It follows that $\Omega$ is a coherent coalition structure. 

It remains to
verify property (\ref{1}). Suppose $x\succeq y$. Then, by construction, the set $U_{(x,y)}\in \Omega$ satisfies
$u(x)\geq u(y)$ for every $u\in U_{(x,y)}$. Conversely, suppose we have
$\lnot(x\succeq y)$. Pick any $U\in\Omega$ so that $U=U_{(x^{\prime},y^{\prime})}$
for some $x^{\prime},y^{\prime}\in X$ with $x^{\prime} \succeq y^{\prime}$. Claim $2$ of \citet{evren2021extension} yields a strictly $\vartriangleright$-increasing function $f:\{x,y,x',y'\}\mapsto \mathbb{R}$ such that $f(x)<f(y)$ and $f(x')\geq f(y')$. In turn, by Corollary 3 in \citet{evren2021extension}, we can extend $f$ to a function $u\in U_{(x^{\prime},y^{\prime})}$ such that $u(x)<u(x)$. Since $U\in\Omega$ is arbitrarily selected, we conclude that we have $x\succeq y$ if and only if there exists a $U\in \Omega$ such that $u(x)\geq u(y)$ for every $u\in U$.

\subsubsection*{(ii) $\implies$ (i)}

Suppose $\succeq$ is represented as in (\ref{1}) for a coalition structure $\Omega$. This clearly implies that $\succeq$ is reflexive.

Assume further that $\Omega$ is coherent, and every $U\in \Omega$ consists of strictly $\vartriangleright$-increasing functions. We shall show that $(\succeq, \succ^*)$ is strictly consistent. Pick any $x,y \in X$ with $x\succ^*y$. Then there exists a $z\in X$ such that either (a) $x\vartriangleright z\succeq y$; or (b) $x\succeq z\vartriangleright y$. In case (a), $z\succeq y$ implies that there exists a $\hat{U}\in\Omega$ such that $u(z)\geq u(y)$ for every $u\in \hat{U}$. Since $x\vartriangleright z$ and the functions in $\hat{U}$ are strictly $\vartriangleright$-increasing, we also have $u(x) > u(y)$ for every $u\in \hat{U}$. Similarly, in case (b), there exists a set $\hat{U}\in \Omega$ such that $u(x) > u(y)$ for every $u\in \hat{U}$.

 In either case,  the existence of such a set $\hat{U}$ and the representation (\ref{1})
jointly imply $x\succeq y$. In fact, we must have $x\succ y$. Indeed, by coherence, any other coalition $V\in\Omega$
contains a function $u\in \hat{U}$ such that $u(y)< u(x)$. Hence, $y\succeq x$ cannot be true by (\ref{1}).

 \bibliographystyle{ecta}
\bibliography{WARPpaper}